\documentclass[preprint]{aastex63}
\usepackage{graphicx}

\newcommand{\kms}{km/s}

\shorttitle{R Aqr}
\shortauthors{Huang et al.}

\begin{document}

\title{Shocks and Photoionization of the Inner 650 AU Jet of the Interacting Binary Star R Aquarii from Multiwavelength \emph{Hubble} Space Telescope Observations}

\correspondingauthor{Caroline Huang}
\email{caroline.huang@cfa.harvard.edu}

\author[0000-0001-6169-8586]{Caroline D. Huang}
\affiliation{Center for Astrophysics $\vert$ Harvard \& Smithsonian \\
60 Garden St.  \\
Cambridge, MA 02138, USA}

\author[0000-0003-1769-9201]{Margarita Karovska}
\affiliation{Center for Astrophysics $\vert$ Harvard \& Smithsonian  \\
60 Garden St.  \\
Cambridge, MA 02138, USA}

\author{Warren Hack}
\affiliation{Space Telescope Science Institute\\
3700 San Martin Drive \\
Baltimore, MD 21218, USA}

\author[0000-0002-7868-1622]{John C. Raymond}
\affiliation{Center for Astrophysics $\vert$ Harvard \& Smithsonian  \\
60 Garden St.  \\
Cambridge, MA 02138, USA}

\author[0000-0002-6752-2909]{Rodolfo Montez Jr.}
\affiliation{Center for Astrophysics $\vert$ Harvard \& Smithsonian  \\
60 Garden St.  \\
Cambridge, MA 02138, USA}

\author[0000-0002-3869-7996]{Vinay L. Kashyap}
\affiliation{Center for Astrophysics $\vert$ Harvard \& Smithsonian  \\
60 Garden St.  \\
Cambridge, MA 02138, USA}

\begin{abstract}

Astrophysical jets are present in a range of environments, including young stellar objects, X-ray binaries, and active galactic nuclei, but their formation is still not fully understood. As one of the nearest symbiotic binary stars, R Aquarii ($D \sim 220$ pc) offers a unique opportunity to study the inner region within $\sim$ 600 AU of the jet source, which is particularly crucial to our understanding of non-relativistic jet formation and origin. We present high-angular resolution ultraviolet and optical imaging from the \emph{Hubble} Space Telescope in six emission-line regions of the inner jet. Using these observations to obtain a range of representative line ratios for our system and kinematic data derived from a comparison with previous studies, we model the shocked gas in order to determine the relative roles of shock heating and photoionization in the R Aquarii system. We find that our shock models suggest a nonzero magnetic field is needed to describe the measured line ratios. We also find that the Mg~II$\lambda\lambda$2795,2802   intensities are overpredicted by our models for most of the jet regions, perhaps because of depletion onto grains or to opacity in these resonance lines.

\end{abstract}

\section{Introduction} \label{sec:intro}

R Aquarii (R Aqr) is a symbiotic binary system, composed of an M6.5-8.5 Mira variable star ($P \sim 387$ d, $M \sim 1-1.75 M_\odot$) and a hot white dwarf (WD, $M \sim 0.5-1.4 M_\odot$) companion, with an orbital period of  $\sim 44$ years \citep{Gromadzki_2009}. The WD is surrounded by a gaseous disk formed by material flowing from the Mira companion. The maximum separation of the binary was measured to be 45.1$\pm0.6$mas, or 9.8 AU in September 2014 by \cite{Schmid_2017} using Very Large Telescope (VLT) imaging.

R Aqr is one of only a small fraction of symbiotics with well-studied jets. Jets are common in a diverse range of astrophysical phenomena, from young stellar objects to active galactic nuclei, but their formation process remains an open question. In most objects, across a range of mass scales, the mass loss rate of the jet has been found to be correlated to the mass accretion rate of the disc \citep{Livio_1997}, thus implying that the mechanisms for jet formation may be common regardless of the type of jet. The study of one class of jets can potentially offer insight into the others. This result requires that jets are produced at the center of accretion disks with a magnetic field and additionally, that powerful jets have a source of energy from the central object. \cite{Huggins_2007} found that jets and tori develop nearly simultaneously in planetary nebulae, with the tori forming slightly earlier than the jets, and suggesting that they are closely related.

Only 10 jets have been detected in about 200 symbiotic stars \citep{Brocksopp_2004}. R Aqr's prominent bioplar jet extends in the NE-SW direction. The jet ejection is generally thought to be related to the orbital position of the binary \citep[e.g.][]{Hollis_1999, Makinen_2004, Liimets_2021}, though some previous observations have suggested otherwise \citep[e.g.][]{Gromadzki_2009}. This system has been detected via a number of spectral features in wavelengths ranging from X-ray to radio \citep[e.g.][]{Schmid_2017, Melnikov_2018, Bujarrabal_2021, Toala_2022}.

At a distance of 218$^{+12}_{-11}$parsecs \citep{Min14}, R Aqr is one of the closest and brightest symbiotic systems, offering a unique opportunity to obtain high spatial resolution and kinematic information from regions near the jet source. It has thus been extensively studied in the hopes that the emission and emission mechanisms in these regions can inform our understanding of the jet formation and origin. 

In this paper, we use the ultraviolet and optical emission and the shock velocity estimated from the evolution of optical features in the system to model the dynamics of the shocked gas and study the relative strengths of photoionization and shocks in the optical line emission. Due to the rapid evolution of the system, simultaneous multiwavelength observations are required in order to understand its complex dynamics. These observations will be used to inform the physical model of the jet formation in subsequent work, which will be based on contemporaneous X-ray, radio, and infrared observations. 

In Section \ref{sec:obs} we describe the observations and analysis. In Section \ref{sec:results} we describe the overall system morphology visible in our emission-line maps and make a comparison with previous observations. In Section \ref{sec:shock_parameters} we discuss the physical implications of the results presented in Section \ref{sec:results}. 

\section{Observations and Data Reduction} \label{sec:obs}

Using the Ultraviolet-Visible (UVIS) channel of the Wide Field Camera 3 (WFC3) instrument on \emph{Hubble Space Telescope} (HST), we obtained narrow-band images of R Aqr with a 40$''\times 40''$ field of view in six emission line regions---Mg~II$\lambda \lambda2795,2802$ (\emph{F280N}), [O~II]$\lambda\lambda3726,2738$ (\emph{F373N}), [O~III]$\lambda 4363$ (\emph{FQ437N}), [O~III]$\lambda$5007 (\emph{F502N}), H$\alpha$ (\emph{F656N}), and [S II]$\lambda$6731 (\emph{F673N})---on 2017-10-13 (GO-14847, PI: Karovska). A detailed summary of these observations and their exposure times can be found in Table \ref{tab:observations}. All of the data presented in this paper were obtained from the Mikulski Archive for Space Telescopes (MAST) at the Space Telescope Science Institute. The specific observations analyzed can be accessed via \dataset[10.17909/j1c4-9h49]{https://doi.org/10.17909/j1c4-9h49}.

We downloaded pipeline-processed images from the HST MAST Archive and generated drizzled and stacked images for each epoch and filter using Drizzlepac 3.1.8. We obtained four subpixel dithered exposures for each filter which were used to create the final products used for the analysis. The subpixel dithering from one exposure to the next for each filter not only allows for identification and removal of cosmic-rays but, more importantly, samples the PSF at a subpixel level. This subpixel sampling enabled drizzling of the images to create final products at the native WFC3/UVIS pixel scale of 0.04$''$/pixel and a pixel scale of 0.025$''$/pixel which we labeled as ‘nominal’ and ‘subpixel’ resolution respectively in the subsequent discussion.

We then aligned each filter’s images with each other in order to obtain line ratios. Our images only contain a single pseudo-point-source – the Mira and white dwarf binary – embedded in extended features that vary in appearance depending on wavelength. The lack of any other point source in the observation prevented the use of conventional point-source alignment code, so we aligned the images using a hybrid alignment method. The alignment was performed with images which were drizzled to a common WCS defined by the exposures taken with the F502N filter. We then identified all extended sources within an approximately 6$''$ $\times$ 6$''$ central region around the binary. Segmentation was performed in the central region to identify and to measure the positions of each of the separate components of the jet, including determining the position of the central binary.

For filters where the binary was measurable, the position of the binary in each filter relative to the position of the binary in the F502N filter’s image was used as the offset between the filters. The binary itself was not visible or distinct enough in the UV filters to support its use in aligning those images with those taken in the visible filters. In those cases, we infer the location of the binary by matching the surrounding extended sources from the jet with the position of the same features seen in the F502N filter. The average offset of the cross-matched sources between the two images was used as the offset of the binary for that filter with respect to the F502N filter. The computed offset for each filter was then used to correct the world coordinate system of each filter’s exposures, then they were drizzled using \emph{AstroDrizzle} to create combined images which were aligned to the F502N filter’s image. This allowed for direct image comparisons of all the drizzled images on a pixel-by-pixel basis across all filters.

\setlength{\tabcolsep}{1em}
\begin{deluxetable}{lcccccc}
\tabletypesize{\scriptsize}
\tablecaption{Summary of \emph{HST} Observations}
\tablewidth{0pt}
\tablehead{\colhead{Dataset} & \colhead{Exposure (s)} & \colhead{Filter} & \colhead{$\lambda_{\text{peak}}$(\AA)} & \colhead{$\lambda_{\text{pivot}}$(\AA)} & \colhead{$\Delta\lambda$(\AA)} & \colhead{Note}
}
\startdata
    IDC501080 &  720 & F275W & 2590.77 & 2710.0 & 480.8 & continuum \\
    IDC501050 & 560 & F280N & 2803.53 & 2796.87 & 22.9  & Mg~II$\lambda\lambda$2795,2802\\
    IDC501030 & 20 & F336W & 3350.96 & 3354.72 & 553.8 & continuum \\
    IDC501BGQ & 240 & F336W & 3350.96 & 3354.72 & 553.8 & continuum \\
    IDC501040 & 400 & F373N & 3740.46 & 3730.10 & 39.2 & [O~II]$\lambda\lambda$3726,3729 \\
    IDC501BHQ & 240 & F373N & 3740.46 & 3730.10 & 39.2 & [O~II]$\lambda\lambda$3726,3729 \\
    IDC501090 & 60 & FQ437N & 4363.63 & 4370.6 & 24.6 & [O~III]$\lambda$4363 \\
    IDC501BQQ & 180 & FQ437N & 4363.63 & 4370.6 & 24.6 & [O~III]$\lambda$4363\\
    IDC501010 & 502 & F502N & 5032.80 & 5009.80 & 57.8 & [O~III]$\lambda$5007 \\
    IDC501020 & 60 & F502N & 5032.80 & 5009.80 & 57.8 & [O~III]$\lambda$5007 \\
    IDC5010A1 & 1.92 & F547M & 5288.47 & 5447.16 & 714.0 & continuum \\
    IDC501BJQ & 60 & F547M & 5288.47 & 5447.16 & 714.0 & continuum \\
    IDC501060 & 80 & F656N & 6556.91 & 6561.1 & 13.9 & $H\alpha$ 6563\\
    IDC501BKQ & 240 & F656N & 6556.91 & 6561.1 & 13.9 & $H\alpha$ 6563\\
    IDC501070 & 80 & F673N & 6757.86 & 6764.5 & 100.5 & [S~II]$\lambda\lambda$6717,6731 \\
    IDC501BLQ & 240 & F673N & 6757.86 & 6764.5 & 100.5 & [S~II]$\lambda\lambda$6717,6731
     \enddata
\tablecomments{All observations were taken on 2017-10-13 using the WFC3/UVIS instrument (GO-14847; PI Karovska). $\Delta \lambda$ is defined as the effective width of the filter---full width at 50\% of peak transmission for wide and medium bands and 10\% of peak transmission for the narrow bands. Exposure times given are the total exposure times. Filter information is obtained from the calibration database system (CDBS) which provides calibration to general observers and the internal \emph{HST} calibration pipeline.}
\label{tab:observations}
\end{deluxetable}

To create fluxed images, we multiplied the drizzled images out{}put by \emph{Astrodrizzle} (in units of $e^-$/s) by the PHOTFLAM keyword in the image header, which results in flux-density images with flux in units of ergs/cm$^2$/s/A/pix.

Emission-line maps created from narrow-band images can contain potential contamination from continuum emission. When available, we use the \emph{HST} wide-band filters to approximate the continuum emission. Concurrent with the narrow-band images, we obtained wide-band images in two bands--\emph{F275W} and \emph{F336W}--as well as one medium-band image--\emph{F547M}. \emph{F275W} and \emph{F336W} are used to model the continuum emission in \emph{F280N} and \emph{F373N} respectively, while \emph{F547M} is used approximate the continuum emission in \emph{FQ437N}, \emph{F502N}, \emph{F656N}, and \emph{F673N}. After aligning and creating flux-density images for all of the filters, we can subtract the wide-band flux density images from the narrow-band flux density images to obtain continuum-subtracted maps for each narrow band. We then convert the subtracted flux density map to a mean flux map by multiplying by the effective width of the narrow-band filter, as given in Table \ref{tab:observations}. The final units of these maps will therefore be in ergs/cm$^2$/s/pix. 

\begin{figure}
    \centering
    \includegraphics[width=1.0\textwidth]{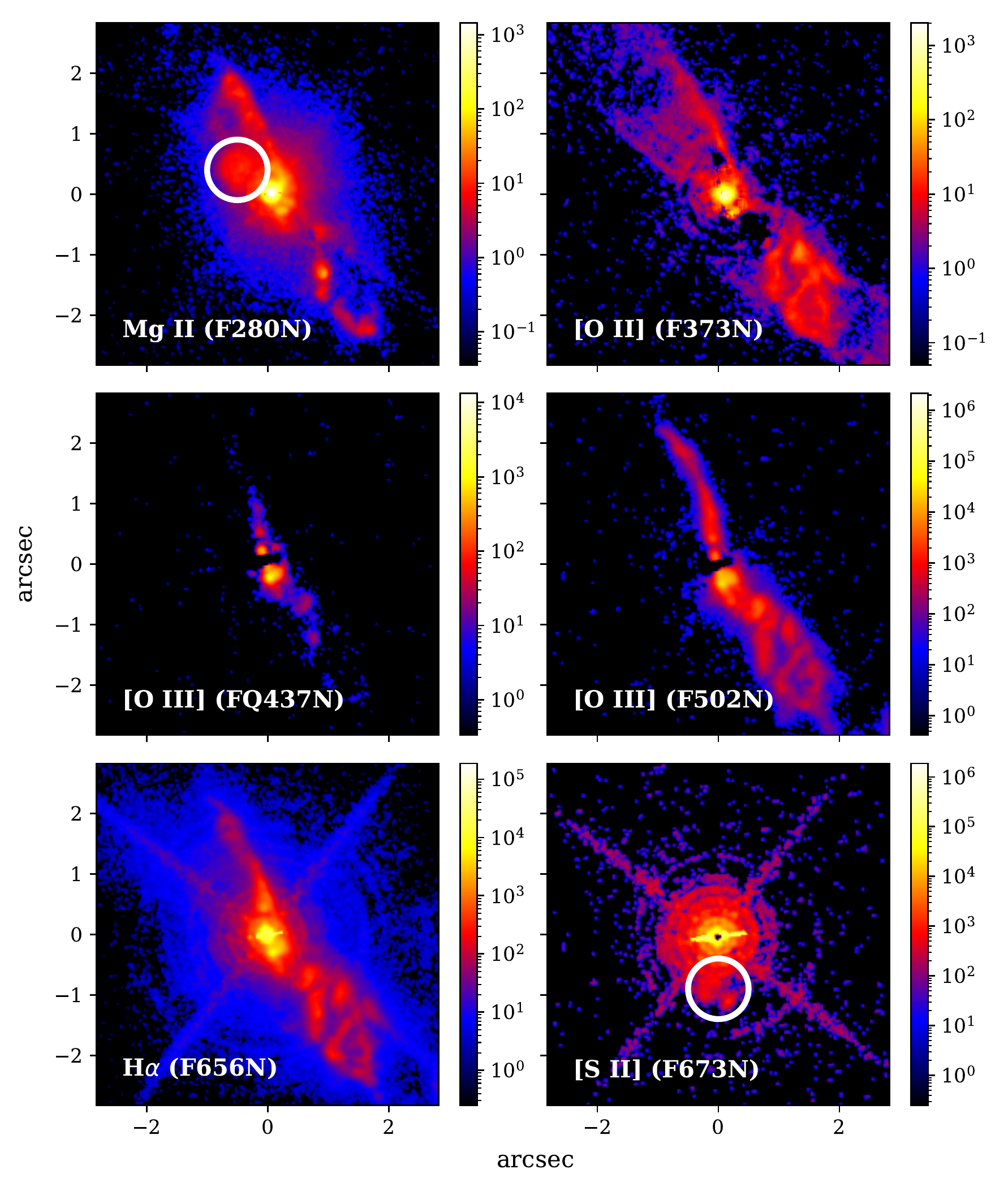}
    \caption{Narrow-band, continuum-subtracted images for the central 1300 $\times 1300$ AU region of the binary, shown in units of ergs/cm$^2$/s/pix. The bandpasses and associated emission lines are given on the bottom left of each map. Low signal-to-noise pixels have been masked (in black). White circles indicate the location of artifacts in \emph{F280N} and \emph{F673N}. Images are oriented with north at the top and east to the left. }
    \label{fig:emissionmaps}
\end{figure}

\begin{figure}
    \centering
    \includegraphics[width=1.0\textwidth]{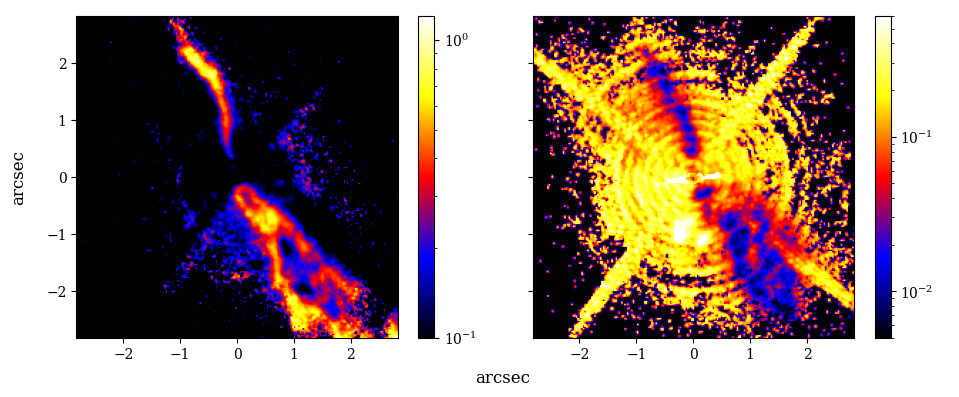}
    \caption{{\bf Left:} 
    Line-ratio map for [O III]/H$\alpha$ with blue indicating a lower line ratio and orange/red indicating a higher line ratio. 
    Regions with [O III]$\lambda$5007 flux below 1.01$\times 10^{-18}$ ergs cm$^{-2}$s$^{-1}$\AA$^{-1}$ or H$\alpha$ flux below 3.14 $\times 10^{-18}$ ergs cm$^{-2}$s$^{-1}$\AA$^{-1}$ were masked due to low signal-to-noise. 
    {\bf Right:} Line-ratio map for [S II]/H$\alpha$. Regions with [S II]$\lambda$6731 flux below 1.01$\times 10^{-18}$ ergs cm$^{-2}$s$^{-1}$\AA$^{-1}$ or H$\alpha$ flux below 3.14 $\times 10^{-18}$ ergs cm$^{-2}$s$^{-1}$\AA$^{-1}$ were masked due to low signal-to-noise.  }
    \label{fig:lineratio}
\end{figure}

Due to the large dynamic range in the images, our shorter-exposure images--appropriate for studying the inner jet--have many pixels of low signal-to-noise outside of the binary and jet. In order to study the morphology, we remove pixels with low signal-to-noise by masking any pixels with counts that do not exceed our minimum source detection threshold. We set this threshold at the background data plus three times the noise. Starting with the counts images, we first estimate the noise by applying a three-sigma clip to the image to remove all sources using the {\tt astropy.stats} module {\tt sigma$\_$clipped$\_$stats} \citep{Astropy_2013,Astropy_2018}. We then use the standard deviation of the remaining background for the ``noise'' and the mean of the remaining background as the average background level. The continuum-subtracted images with these low SNR pixels removed are shown in Figure \ref{fig:emissionmaps}. We create line ratio images by dividing the continuum-subtracted fluxed images. For the line ratios, only pixels where both lines have greater than 3-sigma detection are kept and all others are masked. The ratio images are shown in Figure \ref{fig:lineratio}.

\section{Results}\label{sec:results}

\subsection{System Morphology}

In this paper, we focus on the approximately $1300 \times 1300$ AU ($6 ''\times 6 ''$ assuming $D \sim 218$pc) region around the central binary system. The jet has an overall S-shape, centered on the binary, and extends from northeast to southwest. We identify individual knots as being associated with either the northeast jet or the southwest jet. We begin by briefly describing the morphology of the system in each emission line image shown in Figure \ref{fig:emissionmaps}.

\noindent\emph{F280N}: Mg~II$\lambda\lambda$2795,2802 is an indicator of the ionization state of the gas and is visible in both jets, see Figure \ref{fig:emissionmaps}. In this emission line, one of the most prominent features in the southwest is  the knot SW E1, about 1.5 arcseconds from the central binary. In the northeast jet, there are two main features. The first is emission that appears to originate at the core of the system and arcs about a half arcsecond into the northeast jet, which we have dubbed the ``hinge" (described in greater detail in \ref{sec:fluxes}). The second is emission that traces the northeast jet, extending about two arcseconds from the binary. Finally, there is also a ball of prominent diffuse emission to the northeast of the binary which is an artifact---a known donut ghost signal \citep{Brown_2004}, which is shown in the white circle in Figure \ref{fig:emissionmaps}. 

\noindent\emph{F373N}: Like Mg~II, the [O~II]$\lambda\lambda3726,2738$ doublet (see Figure \ref{fig:emissionmaps}) also provides information on the ionization state of the gas. In particular, it is a density indicator---the line tends to be bright when density is low and is collisionally suppressed at densities above $10^5$ cm$^{-3}$. This emission appears to be emitted in a shell from both jets, outside of the location of the Mg~II$\lambda\lambda$2795,2802 emission. The southwest jet is brighter in [O~II]$\lambda\lambda3726,2738$ relative to the northeast jet. This is in contrast with the Mg~II$\lambda\lambda$2795,2802 emission, for which the reverse was true. 

\noindent\emph{FQ437N and F502N}: The \emph{FQ437N} and \emph{F502N} narrow-band images (see Figure \ref{fig:emissionmaps}) show [O~III] (4363 \AA \ and 5007 \AA \ respectively) emission. The ratio of these lines can serve as an indicator of density and ionization. Many of the features discussed in \ref{sec:fluxes} are prominent in \emph{F502N}. \emph{F502N} emission fills out the central region of the southwest jet and appears to trace the western edge of the northeast jet.  Compared to other line emission, [O~III]$\lambda$5007 in the system is highly collimated and closely follows the jet axis, particularly in the northeast jet. In the southwest jet, [O~III]$\lambda$5007 emission appears to be spread over a larger angle. 


\noindent\emph{F656}: H$\alpha$ emission (6563 \AA, shown in Figure \ref{fig:emissionmaps}) is present in both the northern and southwest jets and is more diffuse than the emission from the other forbidden lines because both collisional excitation and recombination can produce H$\alpha$ emission. Thus, H$\alpha$ emission is useful to compare with the other lines to understand the physical processes occurring at particular location. The H$\alpha$ emission in this system follows a similar distribution as [O~III]. 

\noindent\emph{F673N}: The [S II]$\lambda$6731 (6717/6731 \AA) doublet can also be used to determine gas density. However, the [S~II]$\lambda$6731 maps are saturated because the red Mira component of the central binary is most luminous at longer wavelengths. Thus, this image is dominated by emission from the binary star, R Aqr (shown in Figure \ref{fig:emissionmaps}). The jet structures in this wavelength are difficult to detect against the prominent Airy function of the binary blended PSF. There is a slight excess of diffuse emission at the locations of both jets, which is likely due to the jet rather than the binary. There is also a prominent feature directly to the south of the binary, but this is a known ghost in the \emph{F673N} filter \citep{Brown_2004} and is shown in Figure \ref{fig:emissionmaps} with the white circle. 

\begin{figure}
\centering
  \includegraphics[width=0.49\textwidth]{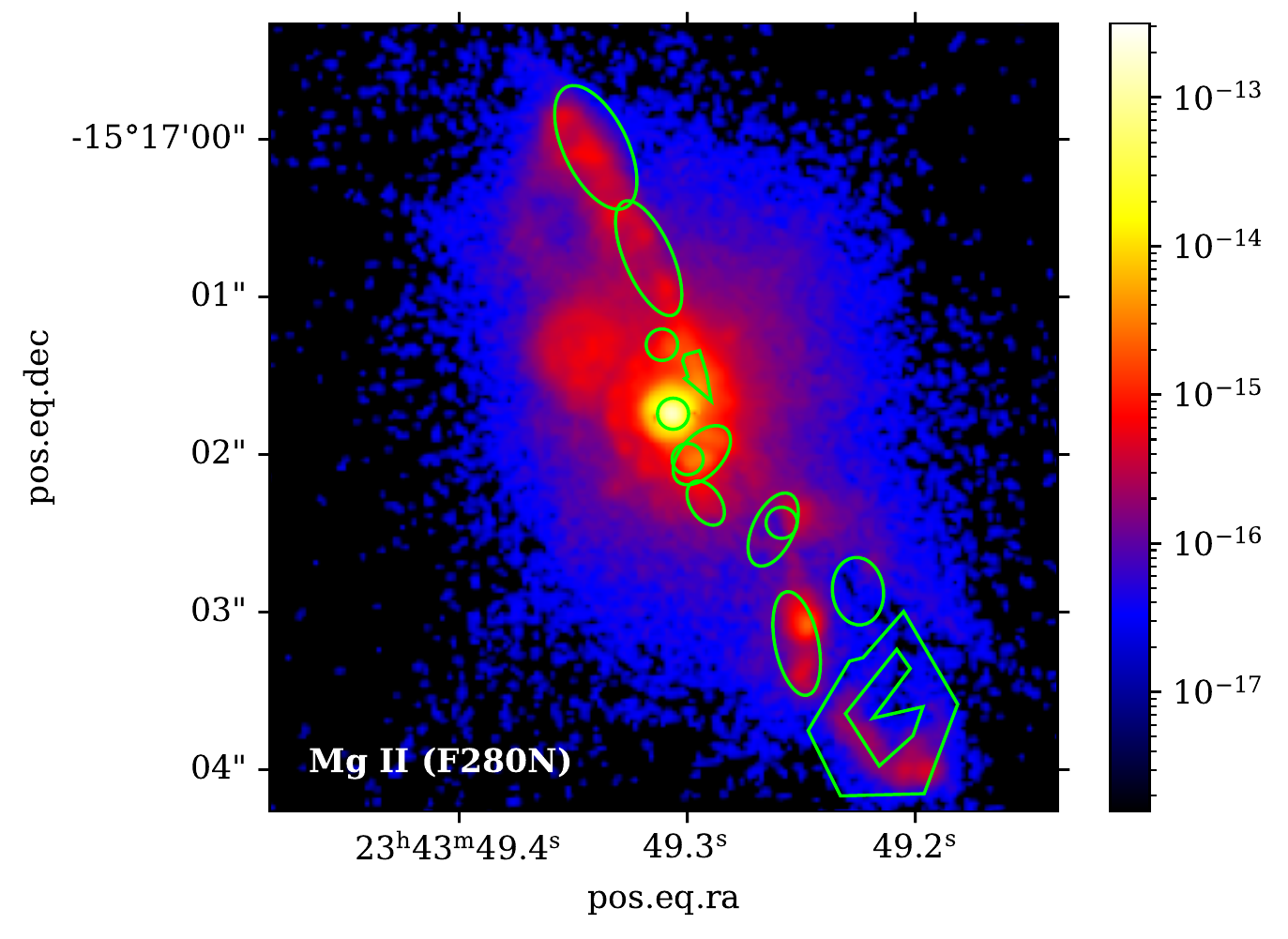}
  \includegraphics[width=0.49\textwidth]{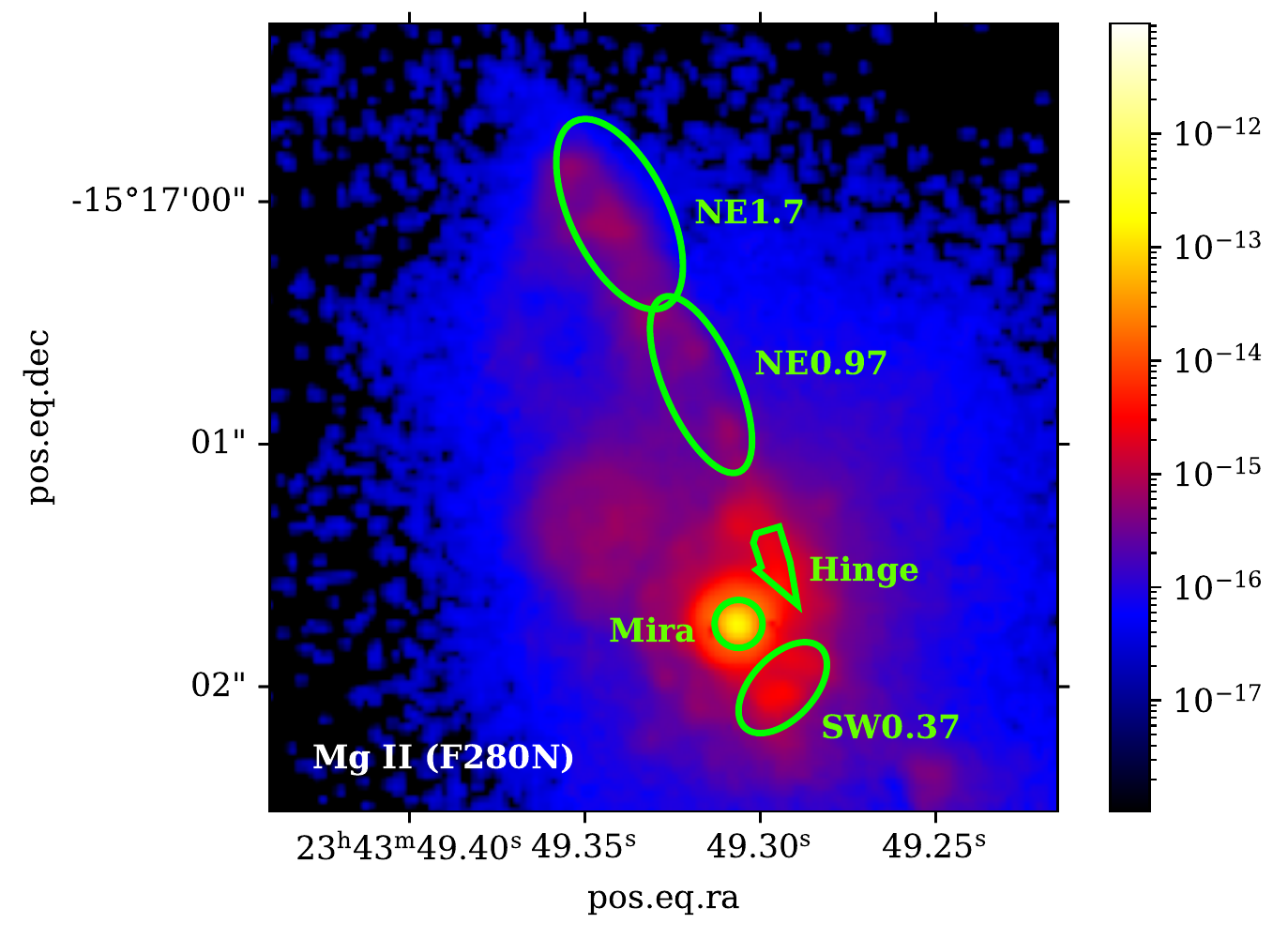}
  \includegraphics[width=0.49\textwidth]{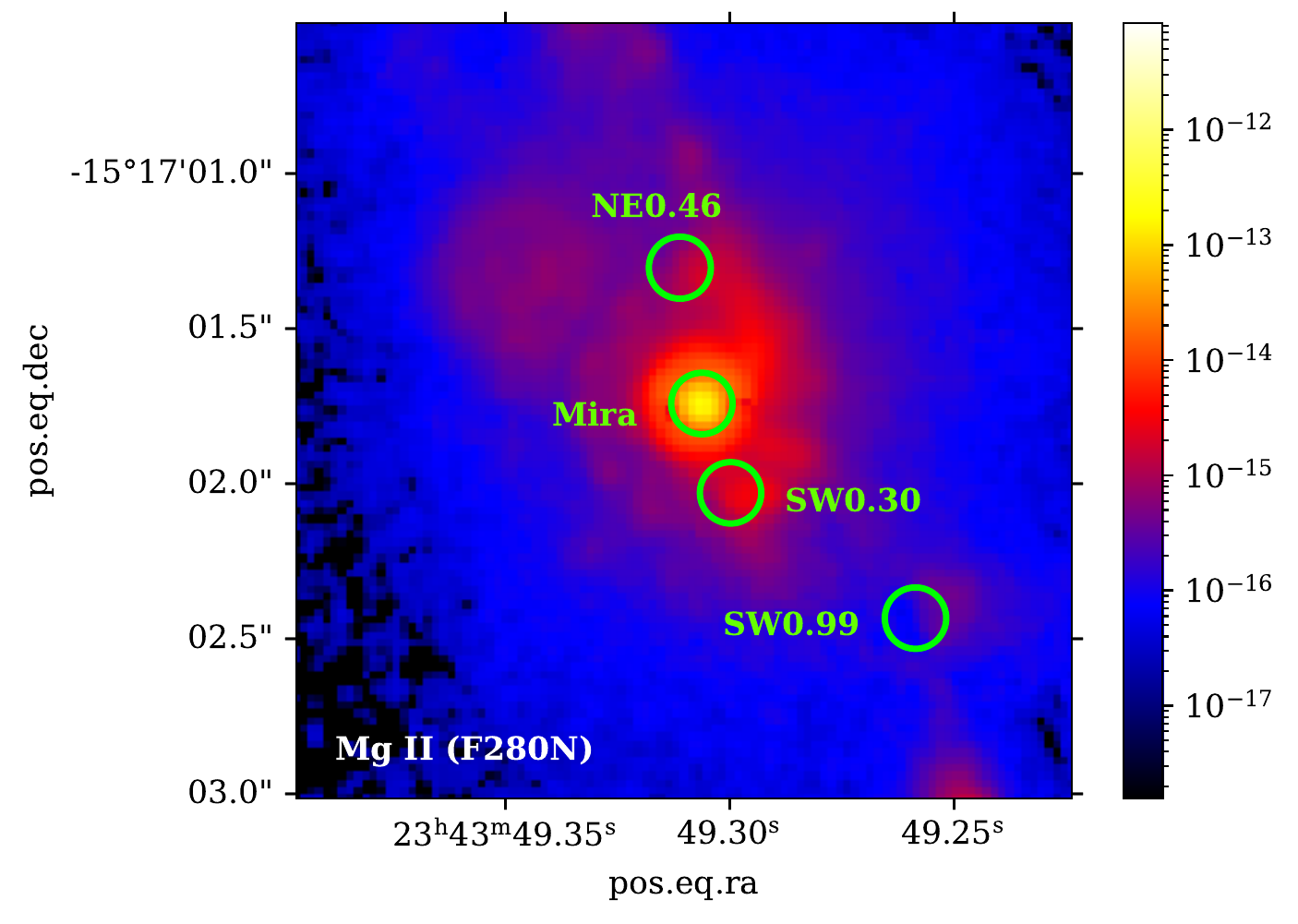} 
  \includegraphics[width=0.49\textwidth]{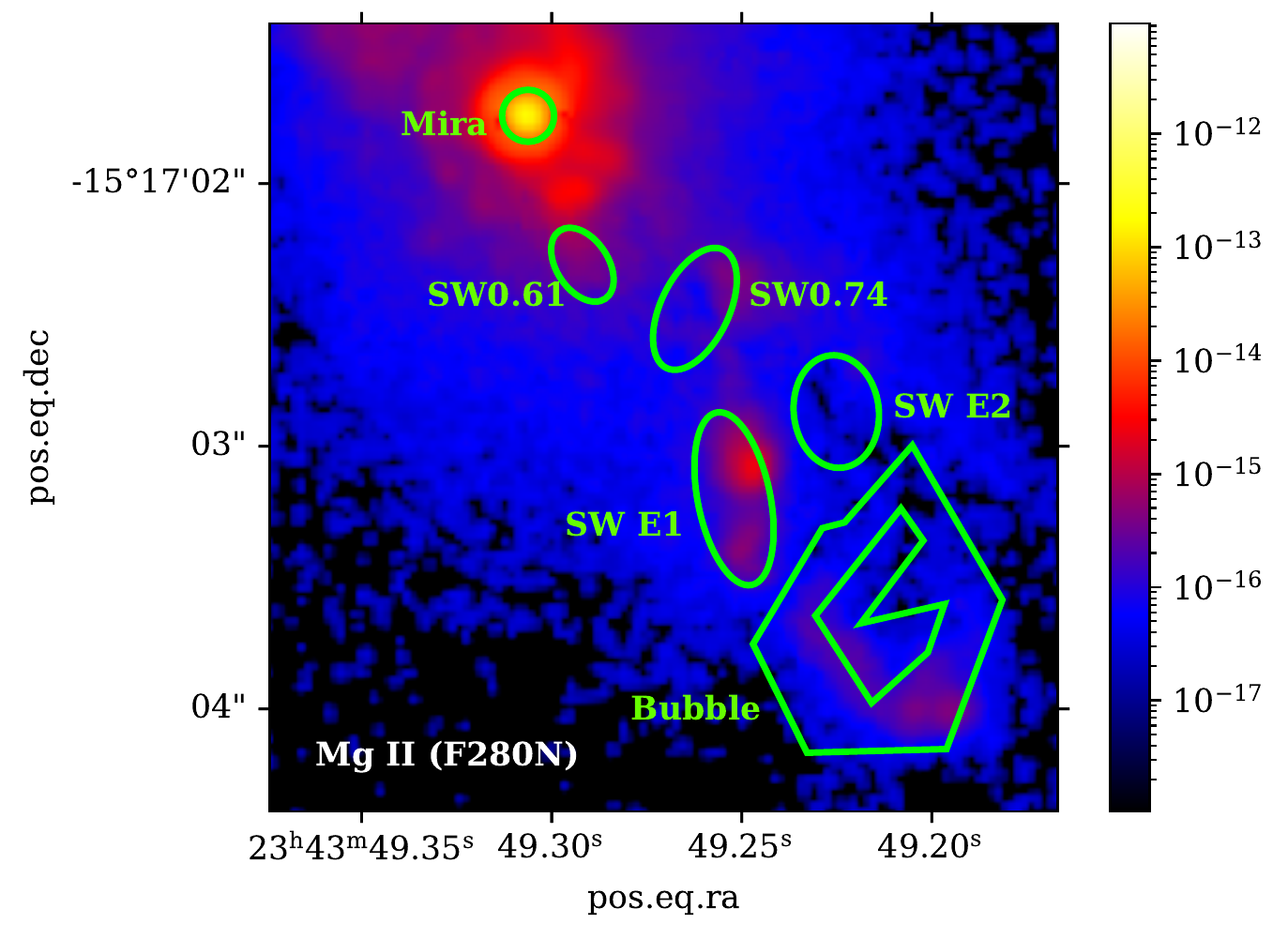}
  \caption{Regions of interest shown in Mg~II$\lambda\lambda$2795,2802 emission maps with orientation and color scheme matching Figure \ref{fig:emissionmaps}. The upper left map shows all of the regions that we studied. The upper right shows the northeast regions, the lower left shows the central regions and the lower right shows the southwest regions. \label{fig:regionsf280n}}
\end{figure}

\begin{figure}
\centering
  \includegraphics[width=0.49\textwidth]{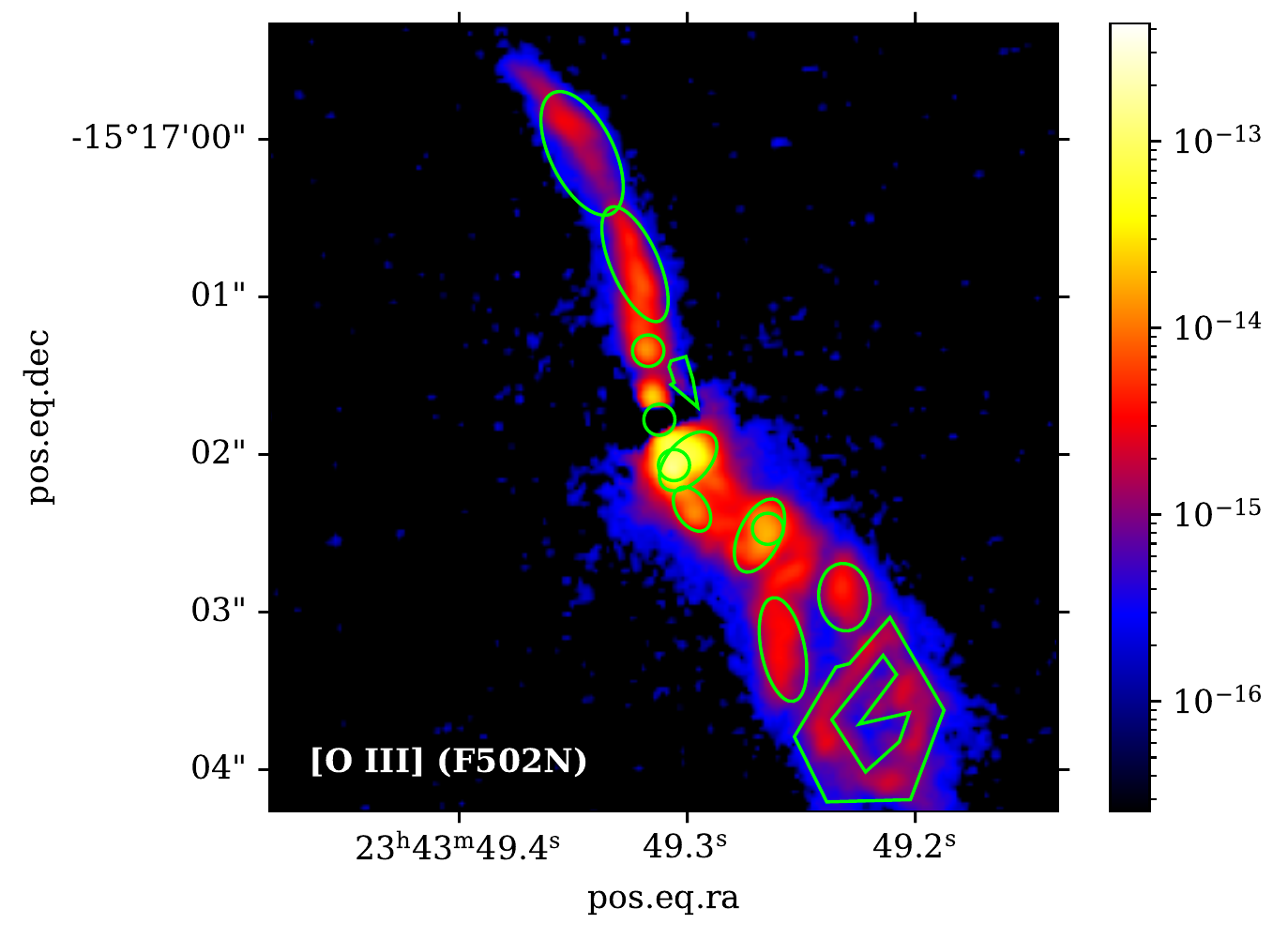}
  \includegraphics[width=0.49\textwidth]{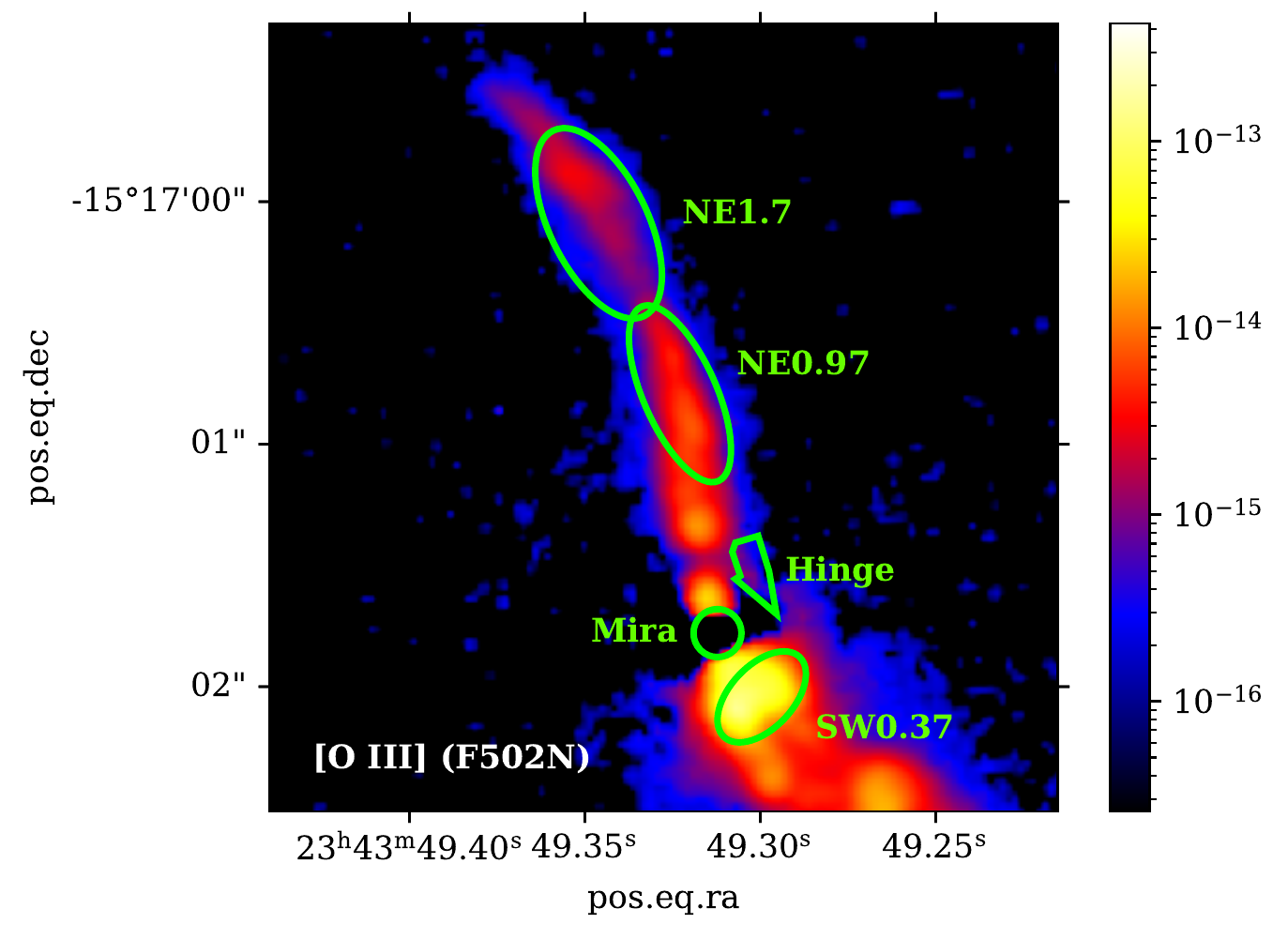}
    \hfill
  \includegraphics[width=0.49\textwidth]{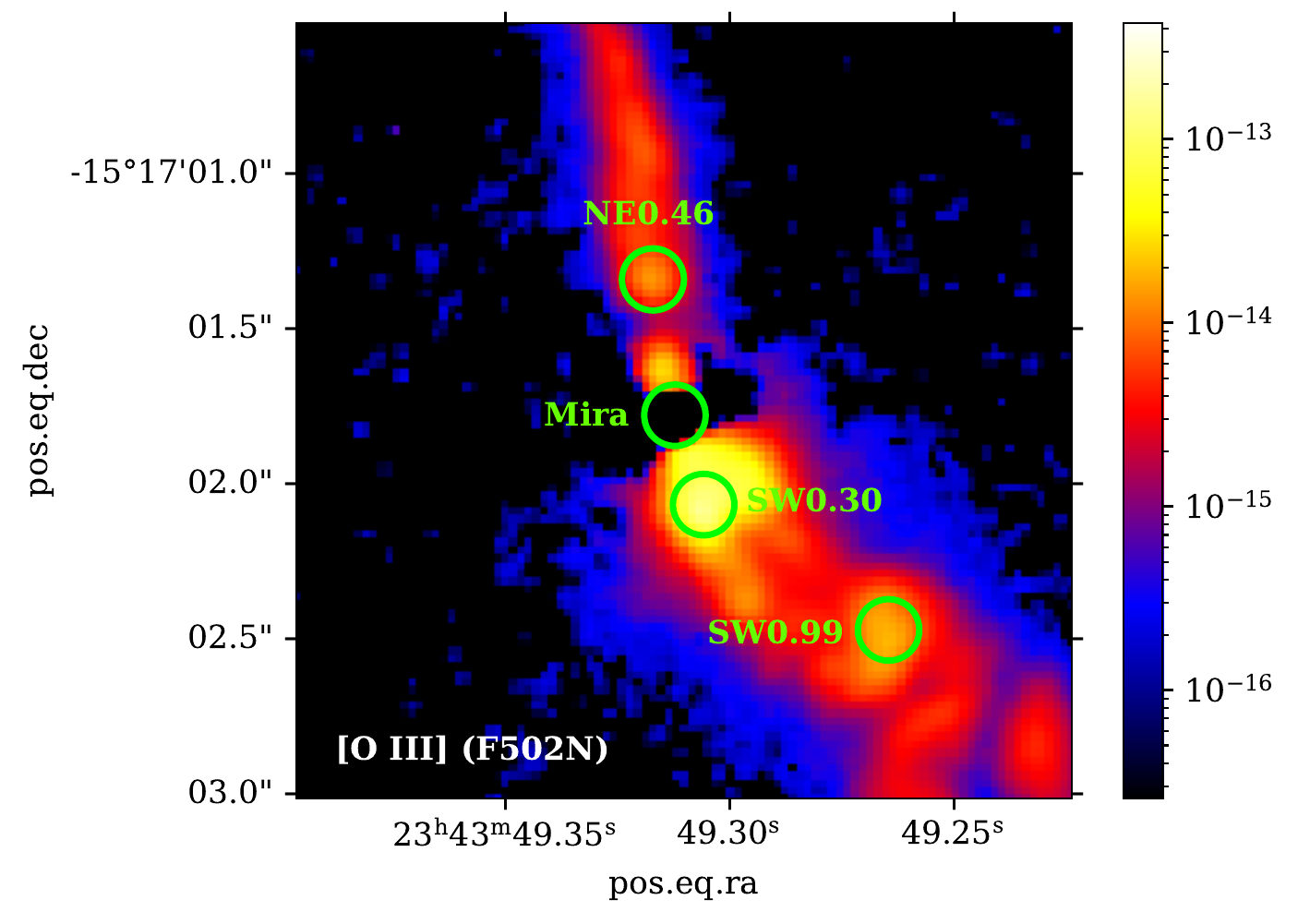}
  \includegraphics[width=0.49\textwidth]{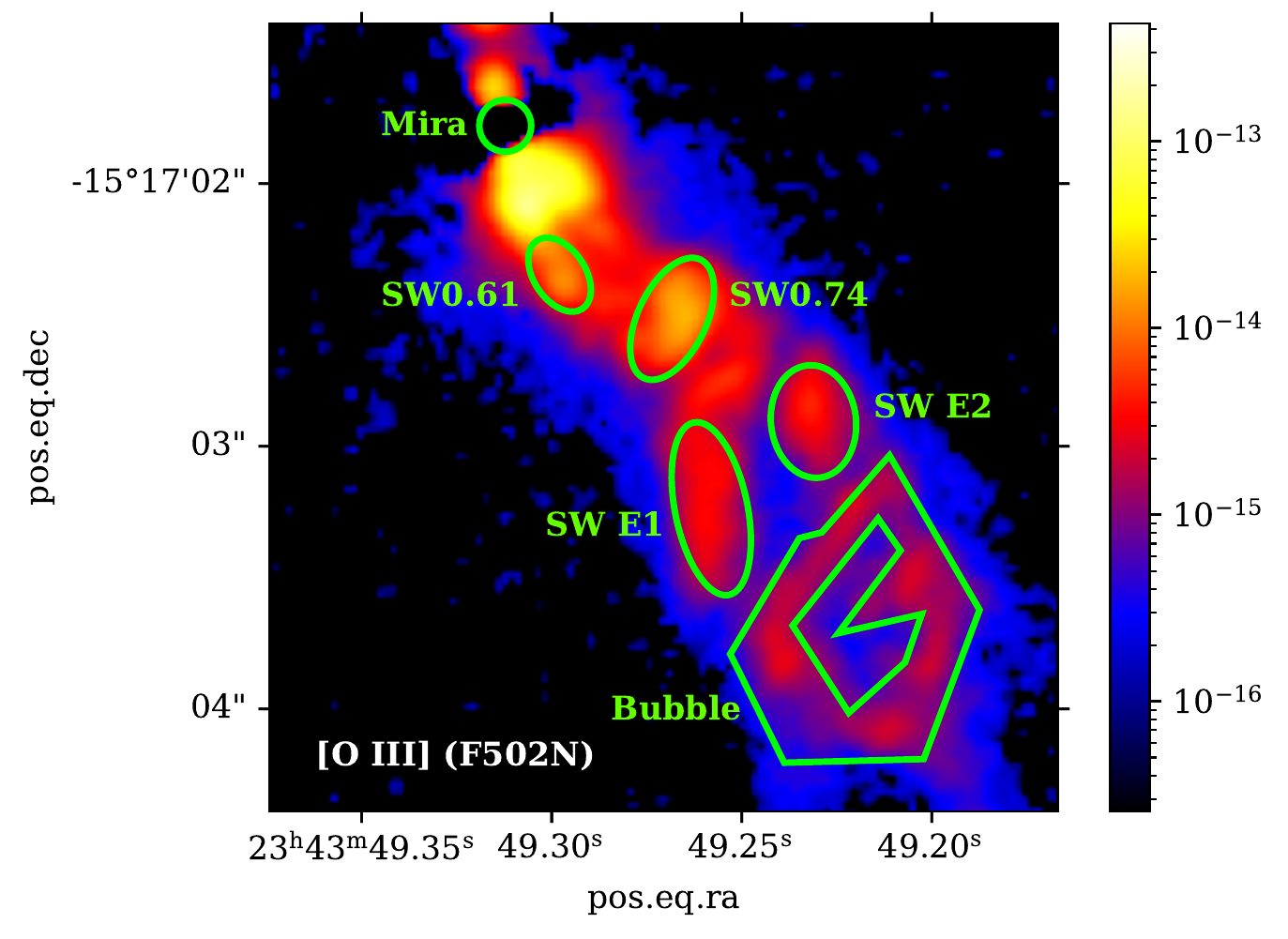}
  \caption{Regions of interest shown in [O~III]$\lambda$5007 emission maps. Orientation and color scheme as in Figure \ref{fig:emissionmaps}. The upper left map shows all of the regions that we studied. The upper right shows the northeast regions, the lower left shows the central regions and the lower right shows the southwest regions. \label{fig:regionsf502n}}
\end{figure}

\begin{figure}
  \centering
   \includegraphics[width=1.0\textwidth]{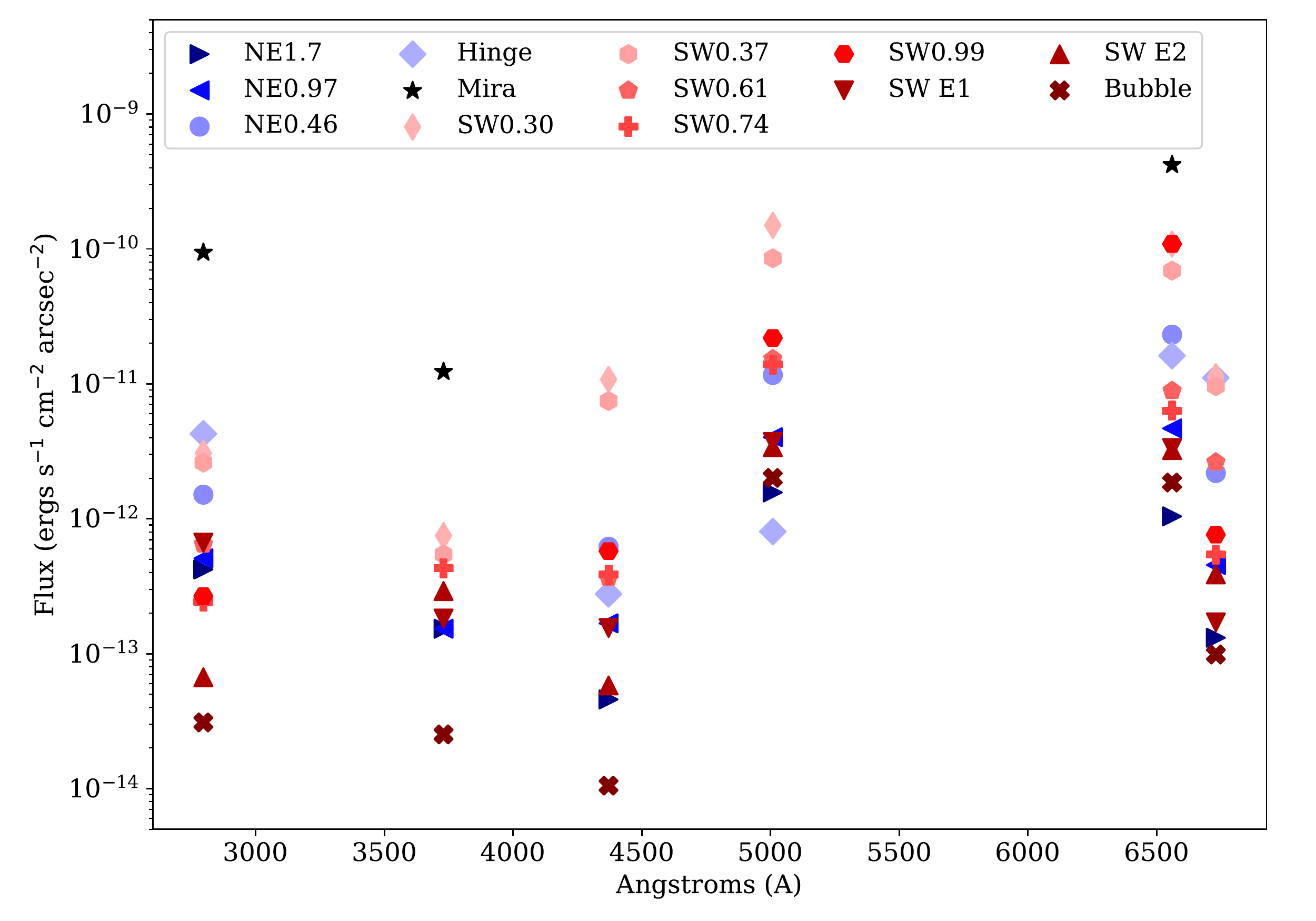}
  \caption{The continuum-subtracted emission line fluxes from the regions discussed in Section \ref{sec:fluxes}\ and shown in Figures \ref{fig:regionsf280n} and \ref{fig:regionsf502n}. Regions are color-coded based on their distance from the central binary, with darker colors corresponding to greater distance. Regions to the north of the binary are shaded in blue, while regions to the south are shaded in red. The Mira is indicated in black.} Flux uncertainties are estimated to be smaller than the size of the points. \label{fig:fluxes}
\end{figure}

\subsection{Fluxes}\label{sec:fluxes}

Based on the morphology seen in the narrow-band images, we select regions (shown in Figures \ref{fig:regionsf280n} and \ref{fig:regionsf502n}) that are potentially useful in understanding the physics of the system and have relatively high S/N. Due to the dynamic nature of the system, the R Aqr system shows significant morphological and brightness changes on the timescale of 10-15 years \citep{Melnikov_2018, Liimets18}. However, whenever possible, we connect the identification of the knots studied in this work with their counterparts identified in previous works, particularly \cite{Schmid_2017} (hereafter S17), which also observed R Aqr in the \emph{F502N} and \emph{F656N} filters and were obtained only three years before the observations in this work. The closest S17 counterparts for each region are listed in the final column of Table \ref{tab:regions}.

We now sum the fluxes in the selected knots using the continuum-subtracted narrow-band emission-line maps and the line-ratio maps. Table \ref{tab:regions} describes the locations of the regions and the dimensions of each aperture we used to obtain fluxes. Table \ref{tab:fluxes} describes the resulting flux measurements for each of these features. The fluxes from Table \ref{tab:fluxes} are shown in Figure \ref{fig:fluxes} and grouped based on their jet location for comparison. 

The northeast jet has an overall conical, collimated shape compared to the southwest jet, which has a lobe-like structure and appears knottier and more diffuse. This difference may be caused by the overall expansion pattern for each jet---for the northeast jet, kinematic data indicates that expansion is roughly north-northeast, whereas for the southwest jet, the expansion is more east-west \citep{Liimets18}, potentially resulting in the wider appearance. Both jets seem to have smaller-scale S-shaped structures that could be the result of precession of the jet. Figure \ref{fig:fluxes} is a comparison of the continuum-subtracted fluxes for every region identified in Tables \ref{tab:regions} and \ref{tab:fluxes}. Most features are named by their quadrant and distance from the central binary, in arcseconds. The exceptions are the ``hinge'' and ``bubble'' features, which are identified by their distinctive morphology and have polygonal apertures. The regions of interest analyzed in this work are not an exhaustive list of all of the knots and interesting features present in the inner jet. Rather, we have chosen a set of regions with representative line ratios for modeling the shock and photoionization in the system. Subsequent papers in other wavelengths, will cover additional regions in the jet such as the bright feature at the start of the northeast jet, between NE0.46 and the central binary. 

\begin{figure}
  \centering
   \includegraphics[width=0.6\textwidth]{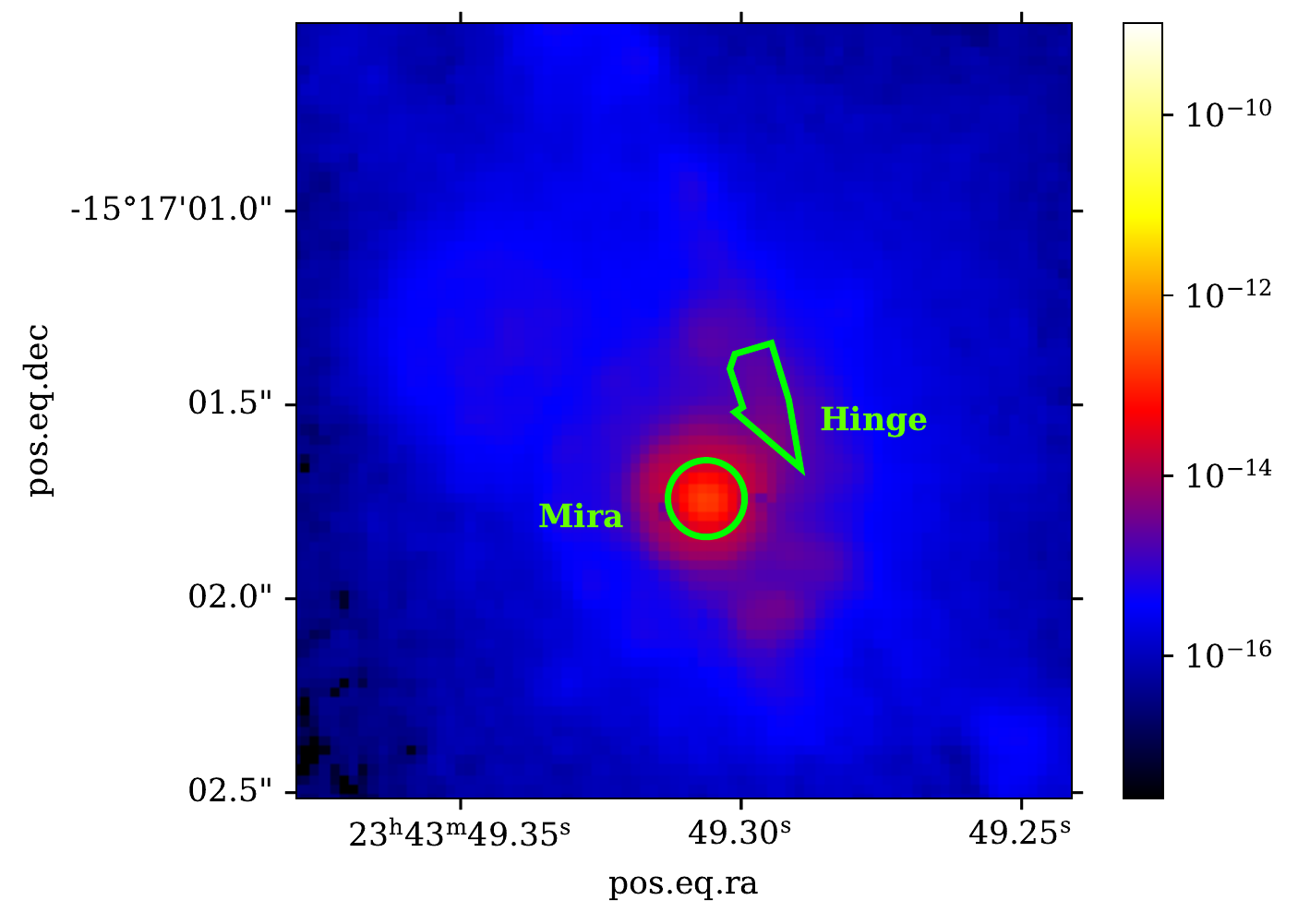}
  \caption{The ``Hinge" region shown in \emph{F280N}/Mg~II$\lambda\lambda2795,2802.$ \label{fig:hinge}}
\end{figure}

\noindent\emph{Hinge:} We dub the emission beginning at the binary and arcing towards the main northeast jet the ``hinge" because its appearance resembles a hinge connecting the start of the northeast jet to the binary. This entire feature is located within half of an arcsecond of the central binary and may show material being transferred to the jet. This is shown in greater detail in Figure \ref{fig:hinge}. The ``hinge" feature is most visible in Mg~II, but also prominent in [O~II]$\lambda \lambda3726-2739$, both [O~III]$\lambda$5007 lines, and H$\alpha$. It was possibly present but difficult to detect in [S~II]$\lambda$6731 due to the brightness of the binary at longer wavelengths. Although the authors did not mention it in their paper, this feature appears to be visible in Figure 7 from S17. Thus, this feature has likely existed as early as 2014.

\noindent\emph{Bubble:} We use the ``Bubble" to refer to large, bubble-shaped region of emission approximately 2 arcseconds to the southwest of the central binary. This was bright in Mg~II, [O~II], [O~III]$\lambda$5007, and H$\alpha$. 

\noindent \emph{Bright knots:} SW0.30 appears to be the origin of the southwest jet and is typically the brightest feature in every wavelength, with the exception of the binary. Overall the southwest jet seems to have more H$\alpha$ emission relative to the northeast jet, which has more Mg~II$\lambda\lambda$2795,2802 and [O~II]$\lambda\lambda3726,2738$ emission. Both jets show strong emission in [O~III]$\lambda$5007. [O~II]$\lambda\lambda3726,2738$ appears to be stronger in the southwest jet relative to the Mg~II$\lambda$2795,2802 emission than is the case in the northeast jet. NE0.46, located within a half arcsecond of the central binary, appears prominently in every wavelength besides [S~II]$\lambda6731$, where it is obscured by the Airy rings of the binary blended PSF. NE0.46 is the other endpoint of the hinge, which originates at the center of the binary. SW0.61, which has a leg-like appearance, did not have detectable [O~II] $\lambda \lambda3726-2739$ emission but was particularly bright in [O~III]$\lambda$5007. SW0.99 is bright in all wavelengths. It has the most emission in [O~III]$\lambda$5007 and H $\alpha$. Similar to SW0.61, while it has [O~III]$\lambda$5007 emission, this knot does not have detectable [O~II] flux $\lambda \lambda3726-2739$.

\setlength{\tabcolsep}{1em}
 \begin{deluxetable}{lcccccccc}
\tabletypesize{\scriptsize}
\tablecaption{Regions of Interest}
\tablewidth{0pt}
\tablehead{\colhead{Name} & \colhead{Dist} & \colhead{Area} & \colhead{$\ell \times w$} & \colhead{P. A.} & \colhead{R.A.} & \colhead{Dec} & \colhead{Shape} & \colhead{S17 Counterpart(s)}
}\startdata
    Mira & 0 & 0.031 & -- & 0 & 23:43:49.5042 & -15:17:04.619 & Circle &  C (jet source)\\ 
    \hline
    \textbf{NE Jet} \\
    Hinge & 0.32 & 0.030 & -- & 334 & 23:43:49.4966 & -15:17:04.325 & Polygon & -- \\
    NE0.46 & 0.46 & 0.031 & -- & 9 & 23:43:49.5066 & -15:17:04.158 & Circle & A$_{\text{N}}$ \\
    NE0.97 & 0.97 & 0.19 & 0.39$''\times 0.15''$ & 9 & 23:43:49.4926 & -15:17:04.895 & Ellipse & A$_{\text{N}}$\\
    NE1.7 & 1.69 & 0.27 & $0.43'' \times 0.20''$ & 16 & 23:43:49.5442 & -15:17:03.026 & Ellipse & C$_{\text{NE}}$ + B$_{\text{NE}}$ \\
    \hline
    \textbf{SW Jet} \\
    SW0.30 & 0.30 & 0.030 & -- & 200 & 23:43:49.4973 & -15:17:04.912 & Circle & A$_{\text{SW}}$\\
    SW0.37 & 0.37 & 0.094 & 0.13$''\times0.23''$ & 215 & 23:43:49.4572 & -15:17:05.353 & Ellipse & B$_{\text{SW}}$ + C$_{\text{SW}}$\\
    SW0.61 & 0.61 & 0.048 & 0.16$'' \times 0.10''$ & 201  & 23:43:49.4889 & -15:17:05.183 & Ellipse & D$_{\text{SW}}$ \\ 
    SW0.74 & 0.74 & 0.10 & 0.25$''\times 0.13''$ & 221 & 23:43:49.5460 & -15:17:02.717 & Ellipse & E$_{\text{SW}}$\\
    SW0.99 & 0.99 & 0.032 & -- & 225 & 23:43:49.4542 & -15:17:05.273 & Circle & E$_{\text{SW}}$\\
    SWE1 & 1.62 & 0.15 & 0.34$''\times 0.14''$ & 209  & 23:43:49.4499 & -15:17:06.062 & Ellipse & F$_{\text{SW}}$ \\
    SWE2 & 1.62 & 0.11 & 0.16$''\times 0.22''$ & 226  & 23:43:49.4214 & -15:17:05.731 & Ellipse & G$_{\text{SW}}$\\
    Bubble & 2.33 & 0.56 & -- & 215 & 23:43:49.4097 & -15:17:06.503 & Polygon & H$_{\text{SW}}$ + I$_{\text{SW}}$ + J$_{\text{SW}}$\\%
    \enddata
\tablecomments{Distance is given in arcseconds from the central binary in \emph{F502N}, with uncertainties of $\pm0.025''$. Aperture areas are given in arcsecond$^2$. $\ell \times w$ refers to the ellipsoid aperture dimensions and is given in arcseconds. P. A. is the position angle of each aperture relative to the location of the central binary. Angles have an uncertainty of $\pm 1^\circ$. The last column gives the name of the closest analog from S17.}
\label{tab:regions}
\end{deluxetable}

\setlength{\tabcolsep}{1em}
 \begin{deluxetable}{lccccccc}
\tabletypesize{\scriptsize}
\tablecaption{Regions of Interest -- Fluxes}
\tablewidth{0pt}
\tablehead{\colhead{Name} & \colhead{F280N-F275W} & \colhead{F373N} & \colhead{FQ437N} & \colhead{F502N} & \colhead{F656N} & \colhead{F673N} & \colhead{F280N-F336W}
}\startdata
Mira & 9.41E-11 & 1.23E-11 & -6.53E-11 & -2.15E-10 & 4.20E-10 & -9.80E-11 & 4.62E-12 \\
\hline 
\textbf{NE Jet} \\
Hinge & 4.27E-12 & -2.31E-14 & 2.77E-13 & 8.05E-13 & 1.62E-11 & 1.11E-11 & 1.96E-13  \\
NE0.46 & 1.51E-12 & -6.31E-14 & 6.21E-13 & 1.16E-11 & 2.32E-11 & 2.18E-12 & 1.85E-13  \\
NE0.97 & 5.10E-13 & 1.53E-13 & 1.68E-13 & 4.02E-12 & 4.68E-12 & 4.55E-13 & 3.58E-14  \\
NE1.7 & 4.21E-13 & 1.53E-13 & 4.58E-14 & 1.57E-12 & 1.04E-12 & 1.31E-13 & 2.65E-14  \\
\hline
\textbf{SW Jet}\\
   SW0.30 & 3.05E-12 & 7.51E-13 & 1.08E-11 & 1.50E-10 & 1.09E-10 & 1.13E-11 & 3.75E-13  \\
SW0.37 & 2.61E-12 & 5.44E-13 & 7.47E-12 & 8.52E-11 & 6.91E-11 & 9.61E-12 & 2.60E-13  \\
SW0.61 & 6.48E-13 & -3.49E-14 & 3.68E-13 & 1.53E-11 & 8.90E-12 & 2.63E-12 & 6.96E-14  \\
SW0.74 & 2.43E-13 & 4.31E-13 & 3.86E-13 & 1.39E-11 & 6.33E-12 & 5.43E-13 & 4.64E-14 \\
   SW0.99 & 2.67E-13 & -3.05E-14 & 5.77E-13 & 2.19E-11 & 1.09E-10 & 7.63E-13 & 6.26E-14  \\
 SWE1 & 6.65E-13 & 1.82E-13 & 1.55E-13 & 3.71E-12 & 3.34E-12 & 1.71E-13 & 4.31E-14  \\
 SWE2 & 6.70E-14 & 2.92E-13 & 5.82E-14 & 3.40E-12 & 3.24E-12 & 3.88E-13 & 1.28E-14  \\
     Bubble & 3.10E-14 & 2.52E-14 & 1.05E-14 & 2.01E-12 & 1.86E-12 & 9.87E-14 & 3.93E-15%
    \enddata
\tablecomments{The continuum-subtracted flux in each of the filters is given in units of ergs/s/cm$^2$/arcsec$^2$. The nearest wide band filter was used to model the continuum for every emission line except for \emph{F280N} for which we used both \emph{F275W} and \emph{F336W} because of the overlap between the \emph{F275W} and \emph{F280N} filters. We did not find significant differences in the line ratios regardless of the continuum filter used.}
\label{tab:fluxes}
\end{deluxetable}

\subsection{Emission Line Ratio Maps}

We use the emission line maps from Figure \ref{fig:emissionmaps} to study the relative intensities of pairs of lines, focusing primarily on the combinations [O~III]$\lambda$5007/H$\alpha$ (\emph{F502N}/\emph{F656N}), and [S~II]$\lambda$6731/H$\alpha$ (\emph{F673N}/\emph{F656N}) (Figure \ref{fig:lineratio}), which are sensitive to the physical parameters of the jet plasma and can provide us with information about ionization state, temperature, and density. [O~III]$\lambda$5007/H$\alpha$ is sensitive to the outflow temperature and can be used to determine the gas temperature within the jet. H$\alpha$ emission can be produced by several different mechanisms and is therefore typically bright in stellar jets. By comparing the distribution and intensity of the other forbidden lines (which typically have a single mechanism of excitation) with H$\alpha$ we can get a better understanding of the mechanisms dominating at different environments in the jet.  

\noindent\emph{[O~III]/H$\alpha$.} Similarly to the [O~III]$\lambda$5007 emission, the line-ratio map (see Figure \ref{fig:lineratio}) for this pair shows that this ratio is highest in the central region of the jet axis. [O~III]$\lambda$5007 emission is highly collimated and causes the opening angle of the jet to appear narrow in this line ratio, which peaks at about 0.7. This shows that the [O~III]/H$\alpha$ line ratio peaks at the gap in between the emission of other forbidden lines and H$\alpha$ in the southwest jet. The overall structure of the [O~III]$\lambda$5007 emission also suggests a helical structure, but the smaller knots and twists in the southwest jet lobe visible in some of the single-line maps are not as prominent. 

\noindent\emph{[SII]$\lambda6731$/H$\alpha$.} The appearance of the [SII]$\lambda$6371/H$\alpha$ line ratio is markedly different from that of [OIII]$\lambda5007$/H$\alpha$. Both \emph{F673N} and \emph{F656N} emission are less collimated than \emph{F502N} emission, causing the line-ratio map to show a wider opening angle for both jets. The complex structure of the southwest jet is more easily visible in this ratio, which peaks at approximately $\sim0.03$. We can also see from the line-ratio map that \emph{F656N} emission fills a wide angle of both jets. Unfortunately, the Mira component of the binary dominates at \emph{F673N}, making it more difficult to see the precise distribution of \emph{F673N} outside of the location of the binary. 

\subsection{Proper Motion Velocities}\label{sec:pmv}

\setlength{\tabcolsep}{1em}
\begin{deluxetable}{cccccccc}
\tabletypesize{\scriptsize}
\tablecaption{Proper Motion Velocities}
\tablewidth{0pt}
\tablehead{\colhead{Region} & \colhead{Velocity (km/s)} & \colhead{$d_{2014} ('')$}  & \colhead{$d_{2017} ('')$} & \colhead{$\sigma_{v}$ (km/s)} & \colhead{$\sigma_d ('')$} & \colhead{P.A. (2014)} & \colhead{P.A. (2017)}
}
\startdata
    A$_{\text{N}}$ & 16 & 0.42 & 0.47 & 8.3 & 0.023 & 3 & 6 \\
    A$_{\text{SW}}$ + B$_{\text{SW}}$ + C$_{\text{SW}}$ & 24 & 0.23 & 0.30 & 9.0 & 0.026 & 230 & 216 \\
    E$_{\text{SW}}$ & 28 & 0.90 & 0.99 & 7.4 & 0.021 & 226 & 225\\
    F$_{\text{SW}}$ & 20 & 1.25 & 1.31 & 7.0 & 0.020 & 218 & 218  \\
    G$_{\text{SW}}$ & 45 & 1.47 & 1.60 & 7.8 & 0.022 & 234 & 229 \\
    Bubble (H$_{\text{SW}}$ + I$_{\text{SW}}$ + J$_{\text{SW}}$) & 114 & 1.97 & 2.30 & 7.2 & 0.021 & 220 & 216
     \enddata
\tablecomments{Proper motion velocities, distances, and uncertainties for the regions defined in Section \ref{sec:pmv}. Proper motions (in km/s) are calculated assuming a constant velocity between the 2014 and 2017 \emph{HST} observations. The distances $d_{2014}$ and $d_{2017}$ (in $''$) given are the distance from the central binary for each region based on the \emph{F502N} 2014 and 2017 observations respectively. P.A. is the position angle, given in degrees, of each knot relative to the central binary in each epoch.
}
\label{tab:pmv_tab}
\end{deluxetable}

\begin{figure}
    \centering
    \includegraphics[width=0.49\textwidth]{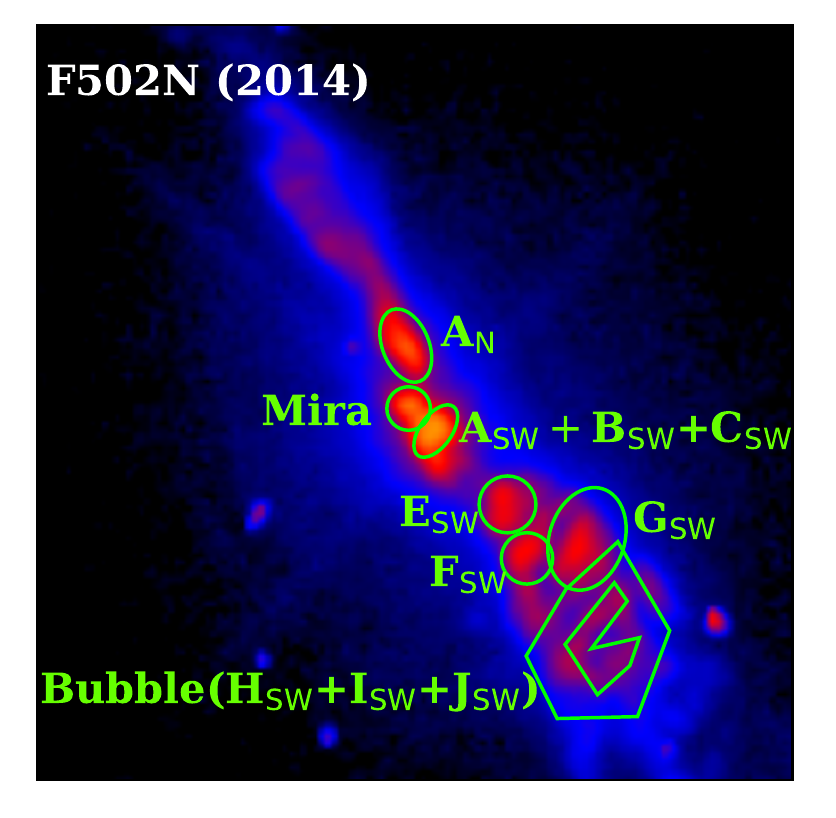}
    \includegraphics[width=0.49\textwidth]{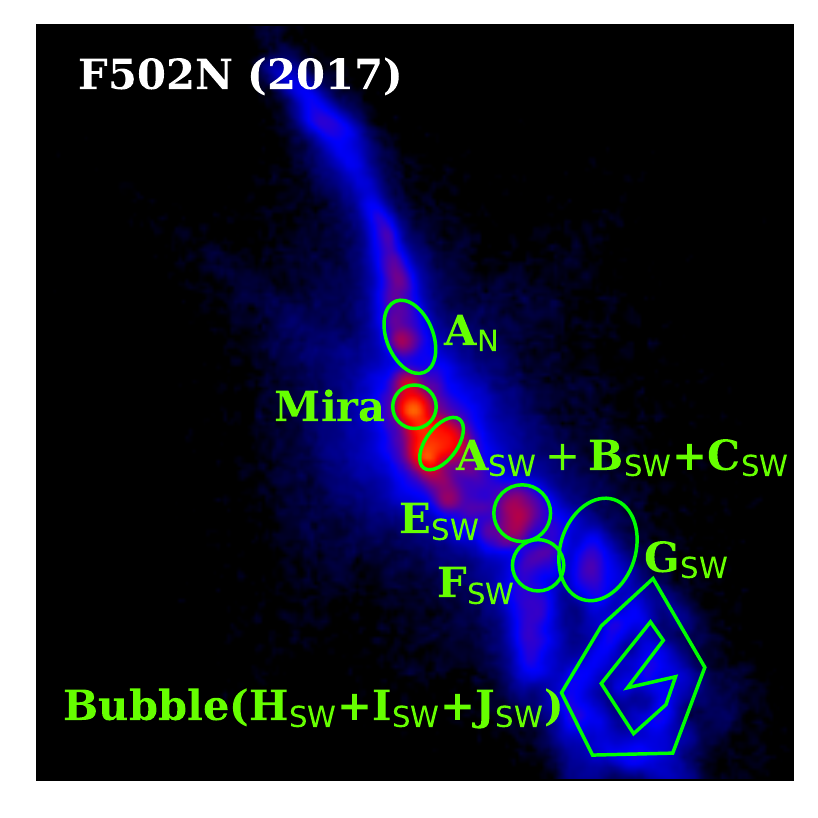}
    \includegraphics[width=0.49\textwidth]{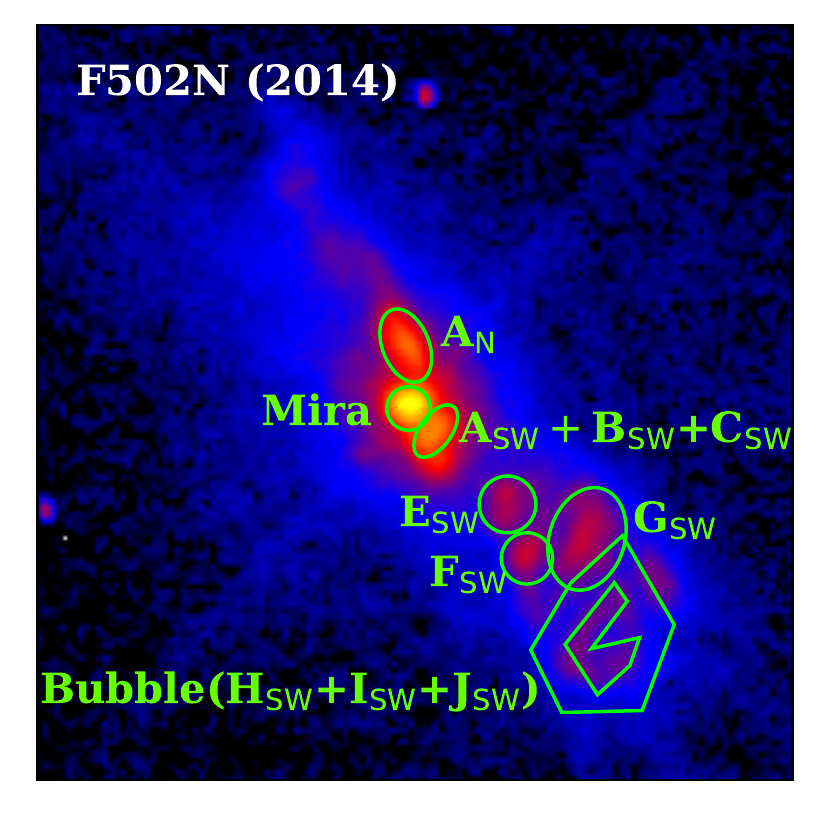}
    \includegraphics[width=0.49\textwidth]{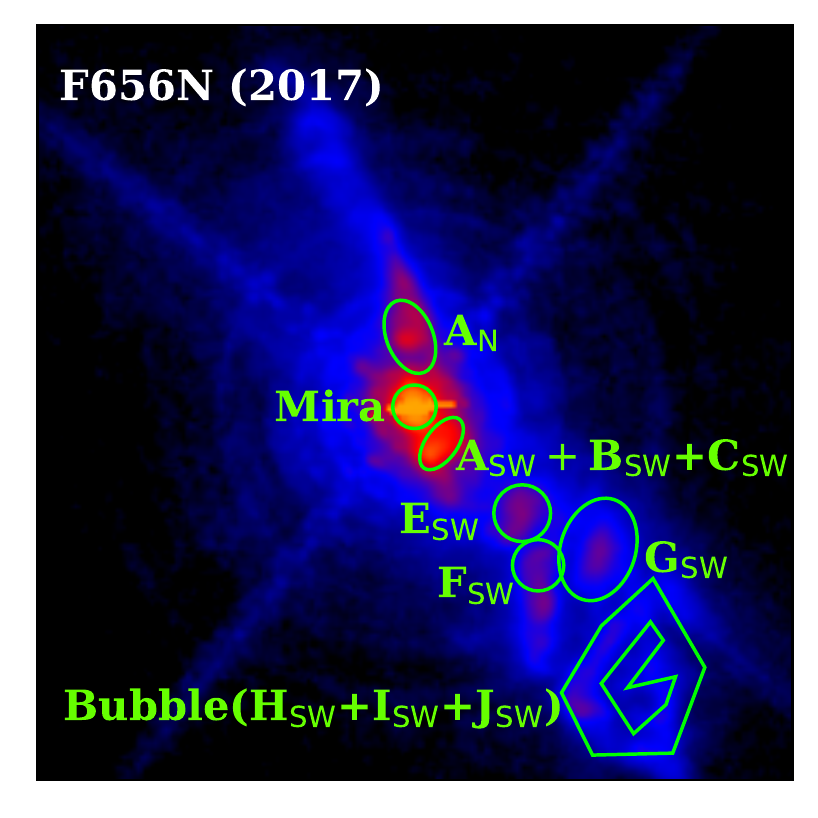}
    \caption{The regions used for comparison of proper motions overplotted on the 2014 HST observations from GO-13339 (PI: Stute) and the 2017 HST observations from GO-14847 (PI: Karovska). Colormaps between both sets of observations have identical scale. }
    \label{fig:schmid_regions}
\end{figure}

The evolution of the system over time can give us insight into the proper motion. This can then be used to rule out various emission mechanisms and provide insight into whether the system is shock or photoionization dominated. While we do not know the exact inclination angle of the system or whether there are standing shocks in the jets, tracking the evolution of the same regions of the jet over time does allow us to place a lower bound on the shock speed.

To measure the proper motions of some of the emission knots in our system, we use archival HST images from 2014 (GO-13339; PI: Stute) in \emph{F502N} and \emph{F656N}. These images had been previously studied in S17 which identified 6 regions in the inner jet. We align these images to ours using the position of the central binary in both bandpasses. 

To conduct as independent as possible of an identification of knots in our observations, the regions of interest in Section \ref{sec:fluxes} are chosen using solely the 2017 data, with apertures based on the 2017 morphology. Due to the evolution of the system, we do not use the apertures defined in previous works to perform the analysis. However, after identifying a set of representative regions of interest, we then cross-match them with their closest analogues from S17 in order to provide an easy comparison between the two lists.

Here, in our proper motion study, we instead use the regions first defined by S17 and shown in Figure \ref{fig:schmid_regions} and Table \ref{tab:pmv_tab}. We only use regions for which we were able to identify similar knots in our observations. While the S17 apertures do not match the morphology of the features in the 2017 data as closely, our primary concern in this case is comparing the distances between the features and the Mira and not measuring fluxes. 
One of the regions previously studied was the binary itself, which we exclude from the proper motion analysis since we measured the proper motions of the other regions relative to the binary. Instead, we add our ``Bubble'' as a sixth region of study along with the other five previously identified in S17. We find small differences ($<5^\circ$) in position angle relative to the Mira for most of the regions between the two sets of observations. The exception is the region closest to the Mira, A$_{\text{SW}} + $B$_{\text{SW}} +$ C$_{\text{SW}}$, which appears to have a change in position angle of $\sim 15^\circ$ degrees. All of these regions and their appearances in both 2014 and 2017 are shown in Figure \ref{fig:schmid_regions}.
 
First, we measure the change in angular separation between each region and the central binary in both epochs. This was converted to a physical distance using the distance to R Aqr of 218 pc from \cite{Min14}. We then assume a constant travel velocity over the 1090 days between the 2014 and 2017 observations in order to calculate the proper motion velocities. The resulting velocities are an average of the velocities obtained for the \emph{F502N} and \emph{F656N} observations for each knots and are given in Table \ref{tab:pmv_tab}. Distances of the knots in the \emph{F502N} filter are given for both epochs. The knots farther from the Mira appear to have a greater velocity, in agreement with the findings from \citep{Melnikov_2018}, which calculated proper motions from a comparison of \emph{HST} observations from 1990-1994. While the authors also obtained more recent observations of R Aqr in 2013 and 2014, the large changes in morphology and brightness of the knots prevented them from identifying more than one common feature between the observations taken in the 2010s and 1990s. The low proper motion velocity for knots close to the jet is also consistent with \citep{Makinen_2004} which found using VLA data that the base of the jet ($\sim 1''$ of the central binary) was moving very slowly. This also suggests that assumption of constant travel velocity for each knot is not correct. 

Given the evolution in the appearance of the knots over time, we include a systematic uncertainty of 1 pixel (0.025$''$) to account for the uncertainty in the location of the knot in between the 2014 and 2017 images. This is added in quadrature with the standard deviation of the measured velocities for each knot. The proper motions we have measured are only the transverse component of the motion. However, the jet inclination angle given by \cite{Hollis_1999} of i$\sim$70 degrees suggests that for all of the knots more central than the Bubble, which have measured proper motion velocities between 20-40 km/s, the knot velocity is $\lesssim 100$ km/s.

\subsection{Flux Error Analysis}\label{sec:errors}

\begin{figure}
\centering
  \includegraphics[width=1.0\textwidth]{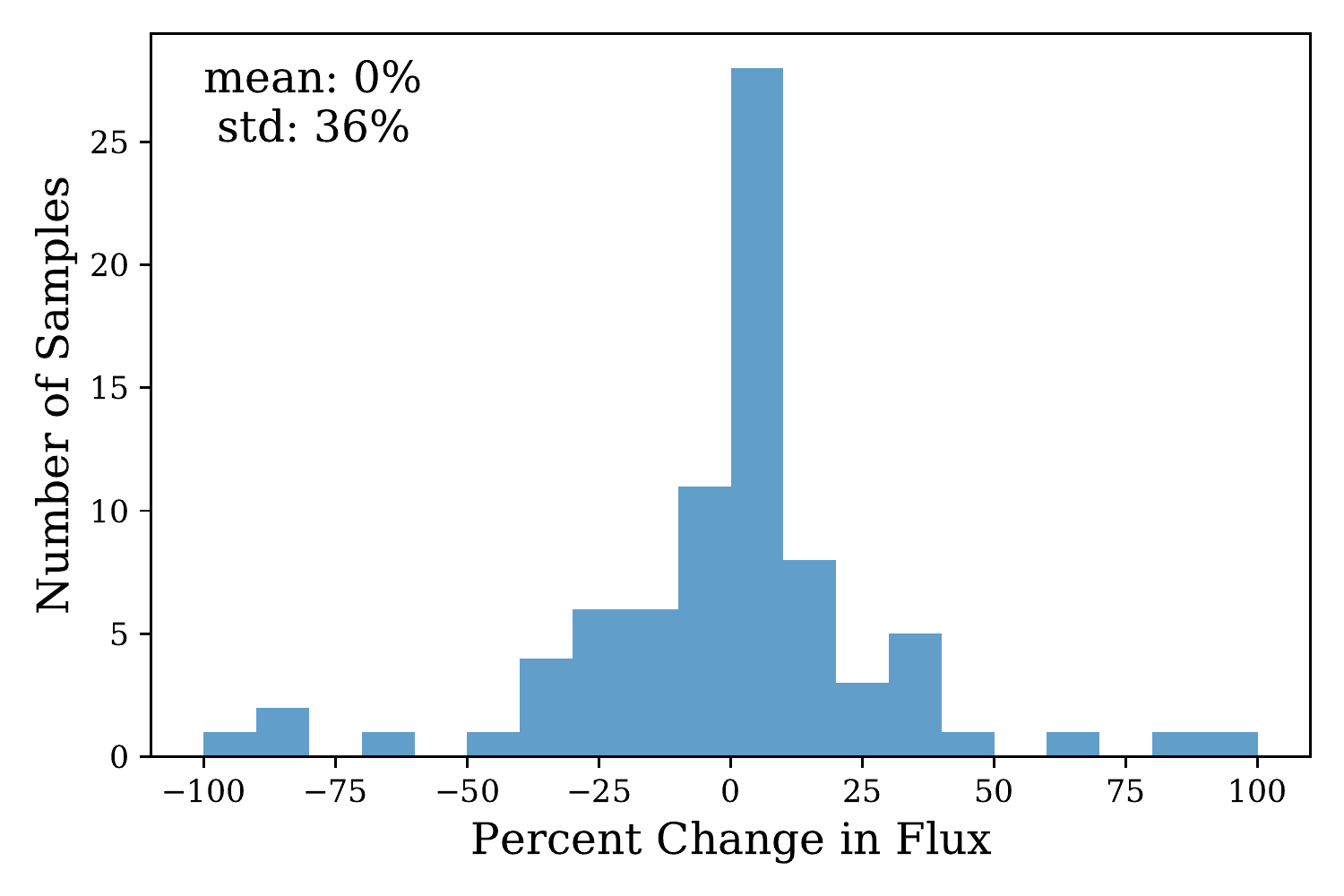}
  \caption{The distribution of changes in flux for the regions in our analysis. The results described in Section \ref{sec:errors} were three-sigma-clipped to remove outliers, but this figure shows the full distribution with no clipping. \label{fig:uncertainties}}
\end{figure}

We are primarily interested in sampling a range of emission line ratios in this system. A complete study analyzing the line ratios of all of the regions in this system is beyond the scope of the paper, and also more complex than our models are able to accommodate. Thus, for simplicity, the regions of study are primarily selected based on their morphology in [O~III]$\lambda$5007 and Mg~II$\lambda\lambda$2795,2802 emission. By centering some regions on [O~III]$\lambda$5007 emission and others on Mg~II$\lambda\lambda$2795,2802 emission we are able to study a representative range of line ratios by comparing regions where different lines dominate. However, choosing regions based on morphology alone also introduces potential biases from the choice of centering each regions. The emission from various lines are not cospatial, and thus the locations of peak surface brightness from each emission knot varies based on the line chosen for identifying the regions. 

In order to estimate the flux uncertainties, we obtain fluxes for the same regions after shifting them by a few pixels to fit the locations of peak surface brightness in other emission lines. We then normalize the fluxes from each region and calculate the percent change in flux compared to the original measurements. We find a distribution shown in Figure \ref{fig:uncertainties} which is sharply peaked around 0\% (i.e., no change in flux), with long tails. After three-sigma-clipping these results to remove a few large outliers, we are left with a standard deviation of 36\%. Thus, we estimate that these measurements are representative for each region to within 36\% flux uncertainty. 

\section{Shock Excitation and Photoionization}\label{sec:shock_parameters}

\subsection{Shock Model Parameters}\label{sec:shockparam}

\begin{figure}
  \centering{}
    \includegraphics[width=1.0\textwidth]{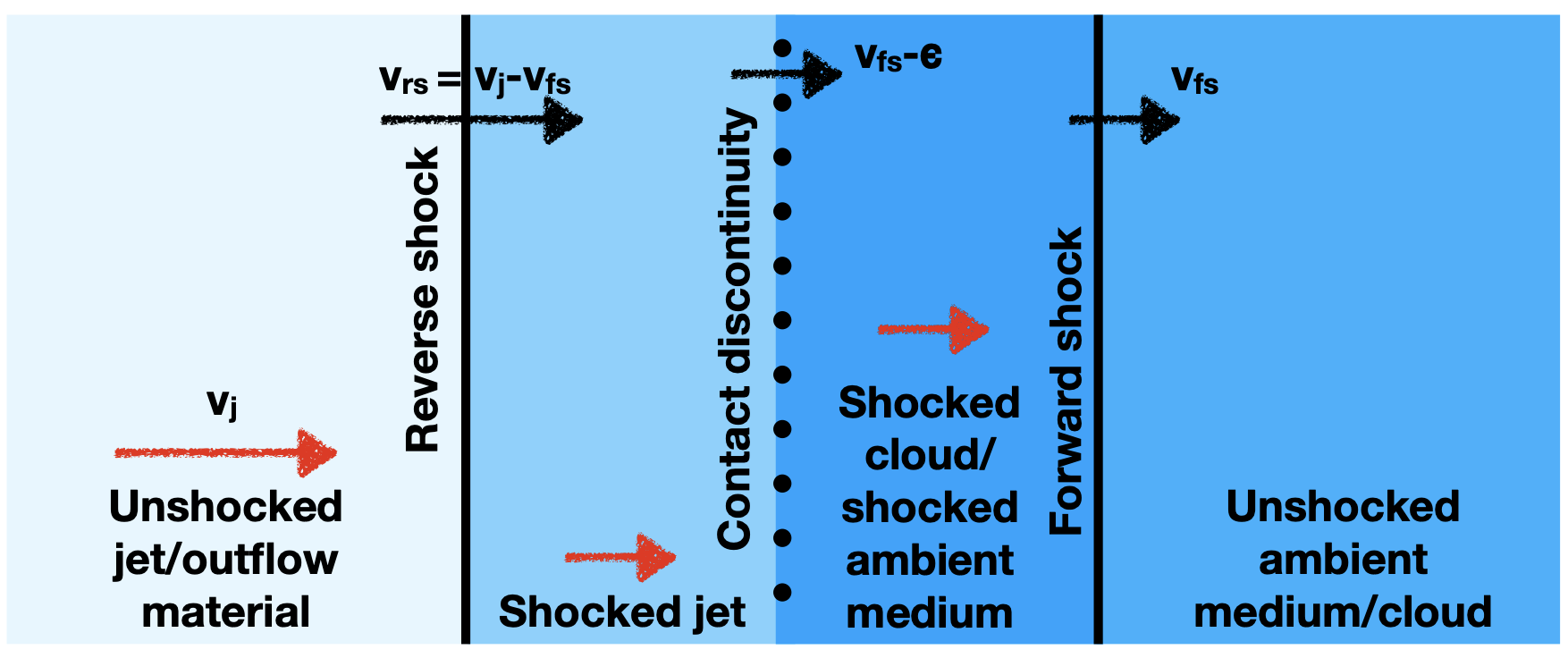}
  \caption{A depiction of the shocks in the emission knots in our system. The lengths of the arrows correspond with the relative magnitudes of the velocities. Red arrows indicate the velocities of the media while the black arrows indicate the velocities of the shocks and contact discontinuity. $v_j$ is the speed of the jet outflow, $v_{rs}$ is the speed of the reverse shock, and $v_{fs}$ is the speed of the forward shock. The shading in each section of the diagram corresponds qualitatively to the density of the material with darker shading indicating denser environs. } 

  \label{fig:shocks}
\end{figure}

To investigate the energetics of the R Aqr flow, one needs to determine the relative roles of shock heating and photoionization.  At greater distances from the jet source, the UV \citep{NicholsSlavin09} and X-ray \citep{Kellogg_2001}, as well as the kinematics \citep{Hollis_2001} demonstrate the importance of shock heating.  \cite{Burgarella_1992} find that shock models without photoionization do not match the relative line fluxes, while models in which the preshock gas is photoionized by the central star can match the observed spectra.  According to \cite{Meier_and_Kafatos_1995} the EUV luminosity of the central source is only $0.05$ L$_\odot$, while \cite{Burgarella_1992} find that $L_{\text{EUV}} = 10$ L$_\odot$ is required.  

Several factors must be kept in mind as we interpret the data.  First, each feature we observe is probably not a single shock, but two shocks: one driven into the ambient gas around the jet and one into the plasma of the jet.  It is also possible that the shocks are formed when faster jet material overtakes slower gas.  In either case, there are likely two shocks whose shock speeds are related by their common ram pressure, $\rho_1 V_1^2 = \rho_2 V_2^2$, so that the ratio of shock speeds is $(\rho_1 / \rho_2)^{1/2}$. Thus the velocity derived from proper motions might apply to a slow shock in the dense Mira wind, while a faster shock in the lower density jet might account for high temperature emission such as [OIII]$\lambda$5007. Second, it should be kept in mind that the shocks may be oblique, in which case only the normal component contributes to the shock heating. An illustration of the shocks is shown in Figure \ref{fig:shocks}.

There are three likely possibilities: 1) Shocks heat the gas and photoionization plays a minor role.  In this case the high temperatures produce relatively strong UV and [O~III] $\lambda$4363 emission; 2) photoionization dominates, and shocks accelerate and compress the gas, but contribute relatively little to the emission.  In this case the UV and [O~III]$\lambda$4363 emission are faint; or 3) photoionization by the central source raises the ionization state of the preshock gas to [O~III]$\lambda$5007 but has relatively little effect on the denser postshock gas.  In this case, the UV will also be faint, and a photoionized precursor may be visible ahead of the shock. In addition, shocks fast enough to strongly heat the gas can sputter dust grains, increasing the Mg abundance, while photoionized gas in the Mira wind is likely to be strongly depleted in Mg.

To assess the feasibility of pure shock heating, we compare the surface brightnesses given in Table 3 with models of shock emission from \citet{Raymond_1979} and \citet{Cox_and_Raymond_1985}, with some updated atomic rates. Model parameters are given in Table \ref{tab:model_params}. Table \ref{tab:hst_obsns} contains the observed line intensities for the brightest R Aqr regions and the intensities predicted by each of the models in Table \ref{tab:model_params}. The line intensities in both Tables are scaled to H$\beta$ = 100, assuming a ratio of H$\alpha$/H$\beta =2.9$. These are steady-flow models of gas that is instantly heated by the shock and cools at nearly constant pressure.  The time-dependent ionization state is computed at each time step including photoionization by EUV radiation produced by the shock.  We assume that the electron and ion temperatures are equal, as is observed for shocks slower than about 500 \kms, and we follow the gas until it cools to 1000 K, where it ceases to produce significant optical radiation.  The significant shock parameters are the shock speed, the preshock density, the perpendicular component of the magnetic field, and the elemental abundances.  We assume the solar abundance set of \cite{Lodders_2009}, or the Lodders abundance set with refractory elements depleted by a factor of two.  Dust depletion can be much more severe than a factor of 2, but we estimate shock speeds around 150 \kms, and shocks at that speed sputter about half of the grain mass and return it to the gas phase \citep{Slavin_2015}.  

{\it Shock Speed:}  The ratio of [O~III]$\lambda$5007 to the Balmer lines is sensitive to the shock speed.  Shocks near 100 \kms \  are too slow to ionize oxygen to [O~III]$\lambda$5007, while shocks approaching 200 \kms \ produce strong photoionizing emission and therefore strong Balmer recombination lines \citep{Hartigan_Raymond_and_Hartmann_1987}.  Shock speeds of about 150 \kms \ produce [O~III]$\lambda$5007 several times as strong as H$\beta$, as is observed in most of the R Aqr features. There is general agreement with \cite{Solf_1992} and \cite{Michalitsianos_1994}.

{\it Preshock densities:}  The surface brightness is proportional to the preshock density, since it is proportional to the mass flux through the shock. The features in R Aqr are not planar shocks observed face-on, and the surface brightness will be geometrically enhanced.  In supernova remnants, the enhancement is often above an order of magnitude, thanks to the large depths of filaments along the line of sight.  For features limited to the width of the R Aqr jet, the enhancement is probably about a factor of 2 or 3.  Thus based on Table \ref{tab:model_params}, pre-shock densities of order $3 \times 10^5~ \rm cm^{-3}$ are required.  The downstream plasma is compressed by a factor of 4 at the shock and a further factor of order 10 as the gas cools.  Densities of order $10^5 ~\rm cm^{-3}$ were inferred by \citet{Meier_and_Kafatos_1995} from the ultraviolet [Si III]/Si III] ratio at larger distances from the binary, in agreement with our estimates. 

{\it Magnetic fields:} A magnetic field can impede the compression of the cooling gas, because the frozen-in field becomes stronger as the shocked gas cools and becomes denser.  At very high densities the [O~II] and [S~II] lines are suppressed, because their critical densities are $10^4 ~\rm cm^{-3}$ and $5\times 10^3 ~\rm cm^{-3}$ respectively.  By comparing Table \ref{tab:model_params} with, for instance, the lower density model grid of \citet{Hartigan_Raymond_and_Hartmann_1987}, it is clear that those lines are suppressed at some level, especially the [O~II] lines.  Those lines are relatively brightest in the features with the lowest surface brightness, and therefore the lowest preshock densities.  None of the models give a very good match to both the [O~II] and [S~II] lines, but a preshock magnetic field of order a few hundred $\mu$G seems to be needed to avoid suppressing the [O~II] and [S~II] lines too strongly.

{\it Elemental abundances:}  The outstanding discrepancy between the models and observations is that the Mg~II line is overpredicted by an order of magnitude.  One possible explanation is that most of the Mg is depleted onto large grains, and that it is not returned to the gas phase as effectively as it is in the models of \citet{Slavin_2015}.  The other is that the optically thick interstellar or Mira wind Mg~II absorption line strongly attenuates the line.  We would need to know the emission line profile in order to estimate the importance of that effect, but it is expected to be strong unless the lines are very strongly Doppler shifted.  The geometry of the emitting knot can also be a factor.  Sheets of emitting gas seen nearly edge-on show severe reduction of resonance lines such as Mg~II \citep{Cornett_1992}.

These results suggest that a model similar to that described in \cite{Burgarella_1992} is responsible for the visible emission, where the shock speeds are 100 km/s in a gas that has been photoionized. While our proper motion velocities are a lower limit, at a jet orientation of order 70$^\circ$, the shock speed is not quite high enough to produce the observed UV and optical fluxes primarily through shock. However, our models differ from the \cite{Burgarella_1992} picture because they assume much higher densities, a range of magnetic fields, and depletion of elements that are likely to be bound in grains. The grain destruction predicted by \cite{Slavin_2015} for shock speeds $\sim 100$ km/s is approximately 40\%. 

The proper motion velocities of $\sim$ 20-40 \kms \ derived in Section \ref{sec:pmv} are too small to account for [O~III]$\lambda$5007 emission if it corresponds to the shock speed in gas that is largely neutral.  Nearly neutral preshock gas would be expected from the low photoionizing fluxes produced by shocks slower than 100 \kms.  However, as mentioned earlier, each emission knot corresponds to a pair of shocks.  The proper motion corresponds to the shock in the ambient or slower material, so the shock in the jet plasma could be faster.  It is also possible that the ambient gas is photoionized by the central source, as suggested by \cite{Burgarella_1992}.  \cite{Meier_and_Kafatos_1995} estimated the temperature of the ionizing flux from the central source  to be about 60,000 K based on the Zanstra method, and they obtained a luminosity of 0.05 L$_\odot$ from a Str\"{o}mgren sphere estimate.  \cite{Burgarella_1992} obtained a somewhat lower temperature of 40,000 K and a luminosity of 10 L$_\odot$  from an analogous Str\"{o}mgren estimate of different features in the jet.  In their picture it appears that the H$\alpha$ flux is mostly due to photoionization, but that the shock compresses the gas and thereby increases the local emissivity.  

The two luminosity estimates span the range of boundary layer luminosities of cataclysmic variables in a high accretion state \citep{Patterson_1985}.  However, the temperatures of those systems are more than twice the temperatures obtained by \cite{Meier_and_Kafatos_1995} and \cite{Burgarella_1992}, both according to theory and from EUV observations (e.g., \cite{Long_1996}, \cite{Mauche_2004}).  Higher temperatures would help with matching the He II fluxes observed in the spectra, though they would make flux from the white dwarf less effective in producing H$\alpha$ emission.  We note that the temperature and luminosity assumed by \citet{Burgarella_1992} would correspond with an emission from a sphere $5\times10^9$ cm in radius. That is much larger than the size of the white dwarf or accretion disk boundary layer, so it seems to be an unlikely combination of parameters. We therefore conclude that shock heating is the main energy source, and that the faster shocks than suggested by the proper motions are likely to be present, but it is also possible that photoionization of the preshock gas, as suggested by \citet{Burgarella_1992}, plays a significant role.  

\subsection{Observation and Model Comparison}\label{sec:comparison}

\begin{figure}
\centering
  \includegraphics[width=1.0\textwidth]{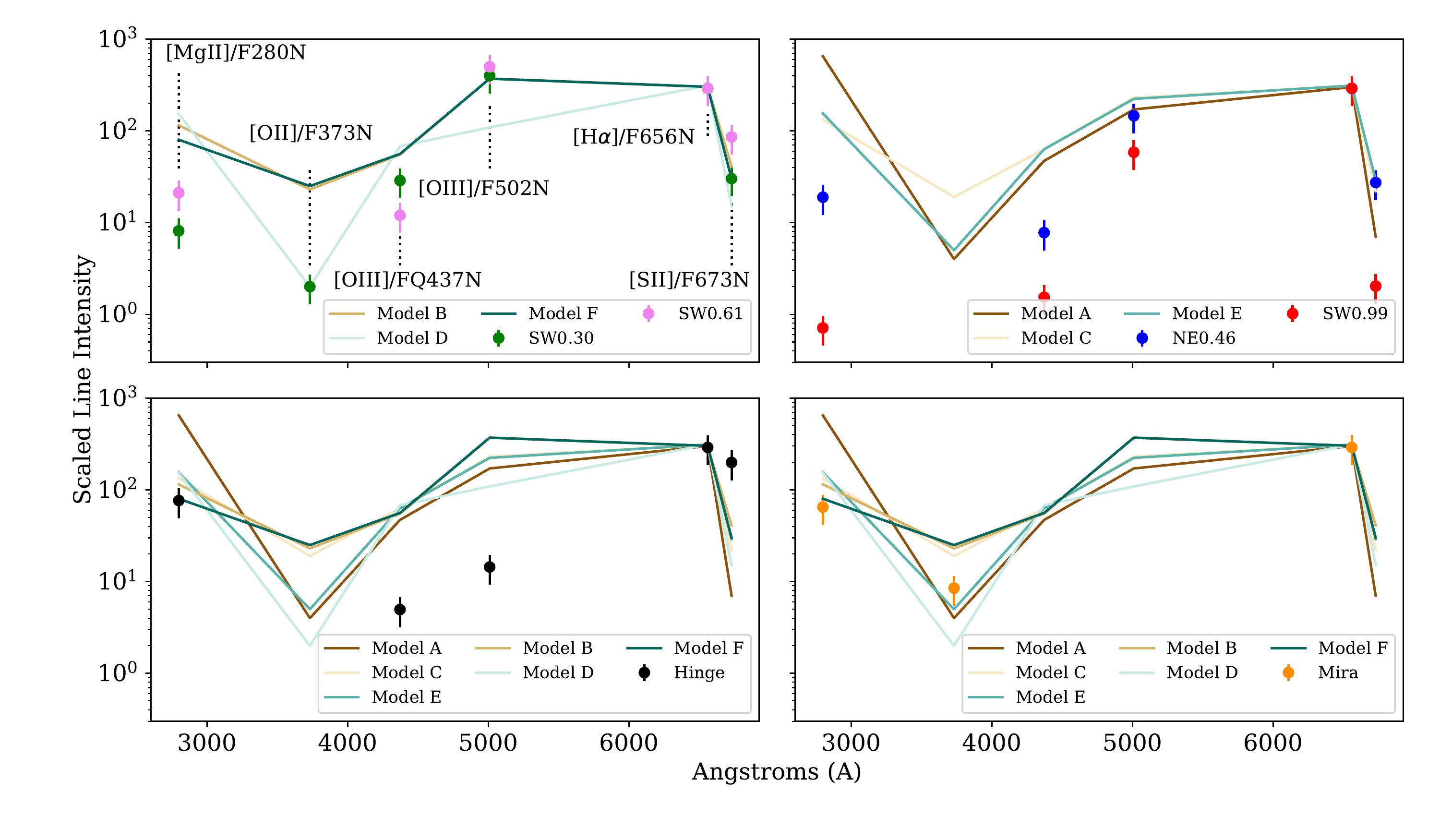}
  \caption{Comparisons of the model fluxes and some selected regions. Both the models and regions are scaled to H$\beta$ = 100. 
  \label{fig:model_v_regions}}
\end{figure}

We now compare the results of the models described in Section \ref{sec:shockparam} with the fluxes obtained in Section \ref{sec:fluxes}. Figure \ref{fig:model_v_regions} shows a comparison of the models and fluxes for a subset of regions with a representative range of line ratios. Line emissions for models and fluxes in Figure \ref{fig:model_v_regions} are scaled to H$\beta$ = 100, assuming H$\alpha$/H$\beta$ = 2.9. Regions are shown with the models that most closely match their emissions with the exception of the Mira/binary and the Hinge, which are both potentially physically distinct from the others. Our models find that some contribution from a magnetic field helps describe the [O~II]$\lambda \lambda3726-2739$ and [O~III]$\lambda$5007 line ratios, which is consistent with the assumption for jet formation made by \cite{Livio_1997}.

While the models were overall able to fit the ratios for the [O~II]$\lambda \lambda3726-2739$ and [O~III]$\lambda$5007 quite well, we found that all of the models overpredicted the amount of Mg~II$\lambda\lambda$2795,2802 emission for every region except the Hinge.  Model F was able to fit the Mg~II$\lambda\lambda$2795,2802 and H$\alpha$ ratio for the Hinge, but overpredicted the emission in the other emission lines. The Hinge region appears morphologically distinct from the other emission knots and the unique line ratio of this region suggests that it is possibly physically distinct from the other regions as well. The flux uncertainties determined in Section \ref{sec:errors} do not appear to have an impact on the interpretation the results.

\section{Discussion and Conclusions}\label{sec:discussion}

While the optical and UV emission of R Aqr's jet far from the Mira has previously been extensively analyzed, \emph{HST} makes it possible to explore the structure and excitation within 1300 AU of the source. The structure at these distances is like that at larger distances, in that it appears to be a narrow jet, and quite likely it is precessing. The precession may be responsible for the changes in position angle relative to the Mira of various knots in the jet between our observations and those from \cite{Schmid_2017}. we estimate proper motion velocities of order $\sim$ 20-40 km/s, which would imply velocities of order 100 km/s when corrected for projection at an inclination angle of 70 degrees \citep{Hollis_1999}. 

We have compared the relative fluxes of the six emission lines we studied with the predictions of shock wave models.  While densities have previously been found to be of the order of 10$^5$ cm$^{-3}$ at distances order 10$''$ from the Mira \citep{Meier_and_Kafatos_1995}, we find that densities an order of magnitude higher are needed to explain the suppression of [O~II]$\lambda\lambda$3726,3729 and [S~II]$\lambda$6731 emission in the knots that we studied.

Previous studies of R Aqr have suggested various mechanisms for the origin of the stellar jet. \citet{Burgarella_1992} considered that the jet could be caused by an expanding, high-velocity stellar wind that is being collimated by interaction with the surrounding medium, while \citet{Meier_and_Kafatos_1995} suggest that the emission is caused by primarily by shocks. \citet{Meier_and_Kafatos_1995} thus assumes lower central source EUV luminosity which does not contribute to optical emission lines. The \cite{Burgarella_1992} model, on the other hand, use a mixture of photoionization and shocks to match the observed relative line intensities.

As a result, one major question that arose from these studies of the R Aqr jet farther from the source was how strongly photoionization and shock heating contributed to the optical and UV line emission. While it is clear that the X-rays come from shocks \citep{Kellogg_2001}, the optical and UV emission was attributed to shock heating by \cite{Meier_and_Kafatos_1995}, while \cite{Burgarella_1992} favored a model in which the shocks moved through gas that had been photoionized by the central source. The uncertainty in the [O~III]$\lambda 4363$ fluxes is large, but for the emission knots within 1$''$ of the source, the relative weakness of the Mg~II and [O~III]$\lambda 4363$ emission at most positions tend to favor low temperatures and/or strong depletion of Mg. Thus those factors suggest a strong role for photoionization.  On the other hand, the proper motions, when corrected for projection effects, indicate speeds that would produce strong enough shocks to produce some [O~III]$\lambda 4363$ emission. Furthermore, the photoinizing source suggested by \cite{Burgarella_1992} seems to be much larger and cooler than those seen in other accreting white dwarfs, suggesting that shock heating may dominate.

Another question is whether the jet is a magnetically-driven jet from near the white dwarf, analogous to many other astrophysical jets, or else an accretion disk wind collimated by its interaction with the dense wind from the Mira. The former picture is favored by \cite{Meier_and_Kafatos_1995}, while the latter is favored by \cite{Burgarella_1992}. 

Our observations do not strongly differentiate between these two pictures, but we estimate that if shock heating dominates the emission, then pre-shock densities above $10^5~\rm cm^{-3}$ are needed to match the observed fluxes, while densities of order $10^6~\rm cm^{-3}$ give the observed degree of suppression for the [O~II] and [S~II] emission. Comparison of those densities indicates a compression factor of 10-20. However, a 100-150 \kms\ shock would compress the gas by factors of 60-200 if it were not moderated by a magnetic field. Thus, our tentative evidence for a field of a few hundred $\mu$G could favor the magnetic jet hypothesis, presented by \citet{Meier_and_Kafatos_1995} though the field could also be generated in a turbulent layer where gas from the Mira wind is entrained in the jet, similar to \citet{Burgarella_1992}. 

In summary, our observations and initial results suggest that UV and optical emission in the R Aquarii system is primarily driven by shock heating from shocks faster than the proper motions of UV and optical features. These results will be explored further in subsequent papers involving both contemporaneous studies in the X-ray and radio and in more recent observations in the UV and optical. 

\begin{deluxetable}{ccccccc}
\tablewidth{0pt}
\tablecaption{Model Parameters}

\tablehead{
\colhead { } &
\colhead {A} &
\colhead {B} &
\colhead {C} &
\colhead {D} &
\colhead {E} &
\colhead {F}
}
\startdata
V         &  150  & 150   & 150   & 150   & 150   & 150 \\
log n$_0$ &  4.5  & 4.0   & 4.5   & 5.0   & 4.5   & 4.5 \\
B$_0$     &  300  & 300   & 300   & 300   &  30   & 1000 \\
He        & 10.93 & 10.93 & 10.93 & 10.93 & 10.93 & 10.93 \\
C         &  8.39 & 7.39  &  7.39 & 7.39  &  7.39 & 7.39 \\
N         &  7.86 & 7.86  &  7.86 & 7.86  &  7.86 & 7.86  \\
O         &  8.73 & 8.73  &  8.73 & 8.73  &  8.73 & 8.73  \\
Ne        &  8.05 & 8.05  &  8.05 & 8.05  &  8.05 & 8.05  \\
Mg        &  7.54 & 6.54  &  6.54 & 6.54  &  6.54 & 6.54  \\
Si        &  7.52 & 6.52  &  6.52 & 6.52  &  6.52 & 6.52  \\
S         &  7.14 & 7.14  &  7.14 & 7.14  &  7.14 & 7.14  \\
Ar        &  6.50 & 6.50  &  6.90 & 6.90  &  6.90 & 6.90  \\
Ca        &  6.33 & 5.33  &  5.33 & 5.33 &  5.33 & 5.33  \\
Fe        &  7.45 & 6.45  &  6.45 & 6.45 & 6.45 & 6.45 %
\enddata

\tablecomments{In each of the models, V is the shock velocity (in km/s), $n_0$ is the preshock density (cm$^{-3}$), and $B_0$ is the preshock magnetic field (in microgauss). We have assumed solar abundances from \cite{Lodders_2009}. Each of the elements is given in relative abundance with H. See Section \ref{sec:shock_parameters} for a detailed discussion of the model parameters. }
\label{tab:model_params}
\end{deluxetable}

\begin{deluxetable}{lrrrrrrrrrrrr}
\tablewidth{0pt}
\tablecaption{Observed Line Intensities, H$\beta$=100}
\tablehead{
\colhead {Line} &
\colhead {SW0.30} &
\colhead {NE0.46} &
\colhead {SW0.99} &
\colhead {SW0.37} &
\colhead {Hinge} &
\colhead {SW0.61} & 
\colhead {A} & 
\colhead{B} & 
\colhead{C} & 
\colhead{D} & 
\colhead{E} & 
\colhead{F}
}
\startdata
Mg II   & 8    & 19  & 1   & 11  & 76  & 21 & 649 & 115 & 133 & 153 & 155 & 80  \\
O II    &  2   &  -  &  -  & 2   & -  & - & 4 & 23 & 19 & 2 & 5 & 25  \\
O III   & 29   & 8   & 2   & 31  & 5  & 22 & 47 & 55 & 63 & 68 & 63 & 56\\
O III   & 399  & 146 & 58  & 358 & 14  & 499 & 171 & 372 & 230 & 109 & 223 & 370 \\
H$\alpha$ & 290 & 290 & 290 & 290 & 290 & 290 & 300 & 300 & 311 & 312 & 311 & 302\\
S II &  27  & 27  & 2  & 40   & 199 & 85 & 7 & 41 & 22 & 15 & 29 & 30 \\
H$\alpha^a$ & 109 & 23 & 108 & 69  & 16  & 9 & 13 & 5 & 15 & 49 & 15 & 12\\
\enddata
\tablecomments{Line intensities (in $1 \times 10^{-12} ~\rm erg~cm^{-2}~s^{-1}~arcsec^2$) relative to H$\beta$ (scaled to 100) in each jet region shown in Figure \ref{fig:model_v_regions}. The last row shows the unscaled H$\alpha$ line intensities. With the exception of SW0.30 and SW0.37, [O~II]$\lambda\lambda3726,3729$ was not detected in our regions.}
\label{tab:hst_obsns}
\end{deluxetable}

\section{Acknowledgments}

We would like to thank the referee for their detailed and helpful comments that have greatly improved the readability of the paper. We would also like to acknowledge Randall Smith and Peter Maksym for their helpful discussions. We are also grateful to AAVSO for the observing campaign and for the R Aqr light curve, critical for planning the HST observations. This work is based on observations made with the NASA/ESA Hubble Space Telescope which is operated by the Association of Universities for Research in Astronomy, Inc., under NASA contract NAS 5-26555. Support for this work was provided by NASA through HST grant GO-14847 from the Space Telescope Science Institute. 



\bibliography{main}{}

\begin{thebibliography}{}
\expandafter\ifx\csname natexlab\endcsname\relax\def\natexlab#1{#1}\fi
\providecommand{\url}[1]{\href{#1}{#1}}
\providecommand{\dodoi}[1]{doi:~\href{http://doi.org/#1}{\nolinkurl{#1}}}
\providecommand{\doeprint}[1]{\href{http://ascl.net/#1}{\nolinkurl{http://ascl.net/#1}}}
\providecommand{\doarXiv}[1]{\href{https://arxiv.org/abs/#1}{\nolinkurl{https://arxiv.org/abs/#1}}}

\bibitem[{{Astropy Collaboration} {et~al.}(2013){Astropy Collaboration},
  {Robitaille}, {Tollerud}, {Greenfield}, {Droettboom}, {Bray}, {Aldcroft},
  {Davis}, {Ginsburg}, {Price-Whelan}, {Kerzendorf}, {Conley}, {Crighton},
  {Barbary}, {Muna}, {Ferguson}, {Grollier}, {Parikh}, {Nair}, {Unther},
  {Deil}, {Woillez}, {Conseil}, {Kramer}, {Turner}, {Singer}, {Fox}, {Weaver},
  {Zabalza}, {Edwards}, {Azalee Bostroem}, {Burke}, {Casey}, {Crawford},
  {Dencheva}, {Ely}, {Jenness}, {Labrie}, {Lim}, {Pierfederici}, {Pontzen},
  {Ptak}, {Refsdal}, {Servillat}, \& {Streicher}}]{Astropy_2013}
{Astropy Collaboration}, {Robitaille}, T.~P., {Tollerud}, E.~J., {et~al.} 2013,
  \aap, 558, A33, \dodoi{10.1051/0004-6361/201322068}

\bibitem[{{Astropy Collaboration} {et~al.}(2018){Astropy Collaboration},
  {Price-Whelan}, {Sip{\H{o}}cz}, {G{\"u}nther}, {Lim}, {Crawford}, {Conseil},
  {Shupe}, {Craig}, {Dencheva}, {Ginsburg}, {Vand erPlas}, {Bradley},
  {P{\'e}rez-Su{\'a}rez}, {de Val-Borro}, {Aldcroft}, {Cruz}, {Robitaille},
  {Tollerud}, {Ardelean}, {Babej}, {Bach}, {Bachetti}, {Bakanov}, {Bamford},
  {Barentsen}, {Barmby}, {Baumbach}, {Berry}, {Biscani}, {Boquien}, {Bostroem},
  {Bouma}, {Brammer}, {Bray}, {Breytenbach}, {Buddelmeijer}, {Burke},
  {Calderone}, {Cano Rodr{\'\i}guez}, {Cara}, {Cardoso}, {Cheedella}, {Copin},
  {Corrales}, {Crichton}, {D'Avella}, {Deil}, {Depagne}, {Dietrich}, {Donath},
  {Droettboom}, {Earl}, {Erben}, {Fabbro}, {Ferreira}, {Finethy}, {Fox},
  {Garrison}, {Gibbons}, {Goldstein}, {Gommers}, {Greco}, {Greenfield},
  {Groener}, {Grollier}, {Hagen}, {Hirst}, {Homeier}, {Horton}, {Hosseinzadeh},
  {Hu}, {Hunkeler}, {Ivezi{\'c}}, {Jain}, {Jenness}, {Kanarek}, {Kendrew},
  {Kern}, {Kerzendorf}, {Khvalko}, {King}, {Kirkby}, {Kulkarni}, {Kumar},
  {Lee}, {Lenz}, {Littlefair}, {Ma}, {Macleod}, {Mastropietro}, {McCully},
  {Montagnac}, {Morris}, {Mueller}, {Mumford}, {Muna}, {Murphy}, {Nelson},
  {Nguyen}, {Ninan}, {N{\"o}the}, {Ogaz}, {Oh}, {Parejko}, {Parley}, {Pascual},
  {Patil}, {Patil}, {Plunkett}, {Prochaska}, {Rastogi}, {Reddy Janga},
  {Sabater}, {Sakurikar}, {Seifert}, {Sherbert}, {Sherwood-Taylor}, {Shih},
  {Sick}, {Silbiger}, {Singanamalla}, {Singer}, {Sladen}, {Sooley},
  {Sornarajah}, {Streicher}, {Teuben}, {Thomas}, {Tremblay}, {Turner},
  {Terr{\'o}n}, {van Kerkwijk}, {de la Vega}, {Watkins}, {Weaver}, {Whitmore},
  {Woillez}, {Zabalza}, \& {Astropy Contributors}}]{Astropy_2018}
{Astropy Collaboration}, {Price-Whelan}, A.~M., {Sip{\H{o}}cz}, B.~M., {et~al.}
  2018, \aj, 156, 123, \dodoi{10.3847/1538-3881/aabc4f}

\bibitem[{{Brocksopp} {et~al.}(2004){Brocksopp}, {Sokoloski}, {Kaiser},
  {Richards}, {Muxlow}, \& {Seymour}}]{Brocksopp_2004}
{Brocksopp}, C., {Sokoloski}, J.~L., {Kaiser}, C., {et~al.} 2004, \mnras, 347,
  430, \dodoi{10.1111/j.1365-2966.2004.07213.x}

\bibitem[{{Brown} \& {Lupie}(2004)}]{Brown_2004}
{Brown}, T.~M., \& {Lupie}, O. 2004, {Filter Ghosts in the WFC3 UVIS Channel},
  Space Telescope WFC Instrument Science Report

\bibitem[{{Bujarrabal} {et~al.}(2021){Bujarrabal}, {Ag{\'u}ndez},
  {G{\'o}mez-Garrido}, {Kim}, {Santander-Garc{\'\i}a}, {Alcolea},
  {Castro-Carrizo}, \& {Miko{\l}ajewska}}]{Bujarrabal_2021}
{Bujarrabal}, V., {Ag{\'u}ndez}, M., {G{\'o}mez-Garrido}, M., {et~al.} 2021,
  \aap, 651, A4, \dodoi{10.1051/0004-6361/202141002}

\bibitem[{{Burgarella} {et~al.}(1992){Burgarella}, {Vogel}, \&
  {Paresce}}]{Burgarella_1992}
{Burgarella}, D., {Vogel}, M., \& {Paresce}, F. 1992, \aap, 262, 83

\bibitem[{{Cornett} {et~al.}(1992){Cornett}, {Jenkins}, {Bohlin}, {Cheng},
  {Gull}, {O'Connell}, {Parker}, {Roberts}, {Smith}, {Smith}, \&
  {Stecher}}]{Cornett_1992}
{Cornett}, R.~H., {Jenkins}, E.~B., {Bohlin}, R.~C., {et~al.} 1992, \apjl, 395,
  L9, \dodoi{10.1086/186476}

\bibitem[{{Cox} \& {Raymond}(1985)}]{Cox_and_Raymond_1985}
{Cox}, D.~P., \& {Raymond}, J.~C. 1985, \apj, 298, 651, \dodoi{10.1086/163649}

\bibitem[{{Gromadzki} \& {Miko{\l}ajewska}(2009)}]{Gromadzki_2009}
{Gromadzki}, M., \& {Miko{\l}ajewska}, J. 2009, \aap, 495, 931,
  \dodoi{10.1051/0004-6361:200810052}

\bibitem[{{Hartigan} {et~al.}(1987){Hartigan}, {Raymond}, \&
  {Hartmann}}]{Hartigan_Raymond_and_Hartmann_1987}
{Hartigan}, P., {Raymond}, J., \& {Hartmann}, L. 1987, \apj, 316, 323,
  \dodoi{10.1086/165204}

\bibitem[{{Hollis} {et~al.}(2001){Hollis}, {Boboltz}, {Pedelty}, {White}, \&
  {Forster}}]{Hollis_2001}
{Hollis}, J.~M., {Boboltz}, D.~A., {Pedelty}, J.~A., {White}, S.~M., \&
  {Forster}, J.~R. 2001, \apjl, 559, L37, \dodoi{10.1086/323667}

\bibitem[{{Hollis} {et~al.}(1999){Hollis}, {Vogel}, {Van Buren}, {Strong},
  {Lyon}, \& {Dorband}}]{Hollis_1999}
{Hollis}, J.~M., {Vogel}, S.~N., {Van Buren}, D., {et~al.} 1999, \apj, 522,
  297, \dodoi{10.1086/307631}

\bibitem[{{Huggins}(2007)}]{Huggins_2007}
{Huggins}, P.~J. 2007, \apj, 663, 342, \dodoi{10.1086/518415}

\bibitem[{Kellogg {et~al.}(2001)Kellogg, Pedelty, \& Lyon}]{Kellogg_2001}
Kellogg, E., Pedelty, J.~A., \& Lyon, R.~G. 2001, The Astrophysical Journal,
  563, L151–L155, \dodoi{10.1086/338594}

\bibitem[{{Liimets} {et~al.}(2021){Liimets}, {Corradi}, {Jones}, {Kolka},
  {Santander-Garcia}, {Sidonio}, \& {Verro}}]{Liimets_2021}
{Liimets}, T., {Corradi}, R.~M.~L., {Jones}, D., {et~al.} 2021, in The Golden
  Age of Cataclysmic Variables and Related Objects V, Vol. 2-7, 41.
\newblock \doarXiv{2003.10753}

\bibitem[{{Liimets} {et~al.}(2018){Liimets}, {Corradi}, {Jones}, {Verro},
  {Santander-Garc{\'\i}a}, {Kolka}, {Sidonio}, {Kankare}, {Kankare}, {Pursimo},
  \& {Wilson}}]{Liimets18}
{Liimets}, T., {Corradi}, R.~L.~M., {Jones}, D., {et~al.} 2018, \aap, 612,
  A118, \dodoi{10.1051/0004-6361/201732073}

\bibitem[{{Livio}(1997)}]{Livio_1997}
{Livio}, M. 1997, in Astronomical Society of the Pacific Conference Series,
  Vol. 121, IAU Colloq. 163: Accretion Phenomena and Related Outflows, ed.
  D.~T. {Wickramasinghe}, G.~V. {Bicknell}, \& L.~{Ferrario}, 845

\bibitem[{{Lodders} {et~al.}(2009){Lodders}, {Palme}, \& {Gail}}]{Lodders_2009}
{Lodders}, K., {Palme}, H., \& {Gail}, H.~P. 2009, Landolt B\&ouml;rnstein, 4B,
  712, \dodoi{10.1007/978-3-540-88055-4\_34}

\bibitem[{{Long} {et~al.}(1996){Long}, {Mauche}, {Raymond}, {Szkody}, \&
  {Mattei}}]{Long_1996}
{Long}, K.~S., {Mauche}, C.~W., {Raymond}, J.~C., {Szkody}, P., \& {Mattei},
  J.~A. 1996, \apj, 469, 841, \dodoi{10.1086/177832}

\bibitem[{{M{\"a}kinen} {et~al.}(2004){M{\"a}kinen}, {Lehto}, {Vainio}, \&
  {Johnson}}]{Makinen_2004}
{M{\"a}kinen}, K., {Lehto}, H.~J., {Vainio}, R., \& {Johnson}, D.~R.~H. 2004,
  \aap, 424, 157, \dodoi{10.1051/0004-6361:20035866}

\bibitem[{{Mauche}(2004)}]{Mauche_2004}
{Mauche}, C.~W. 2004, \apj, 610, 422, \dodoi{10.1086/421438}

\bibitem[{{Meier} \& {Kafatos}(1995)}]{Meier_and_Kafatos_1995}
{Meier}, S.~R., \& {Kafatos}, M. 1995, \apj, 451, 359, \dodoi{10.1086/176225}

\bibitem[{{Melnikov} {et~al.}(2018){Melnikov}, {Stute}, \&
  {Eisl{\"o}ffel}}]{Melnikov_2018}
{Melnikov}, S., {Stute}, M., \& {Eisl{\"o}ffel}, J. 2018, \aap, 612, A77,
  \dodoi{10.1051/0004-6361/201731749}

\bibitem[{{Michalitsianos} {et~al.}(1994){Michalitsianos}, {Perez}, \&
  {Kafatos}}]{Michalitsianos_1994}
{Michalitsianos}, A.~G., {Perez}, M., \& {Kafatos}, M. 1994, \apj, 423, 441,
  \dodoi{10.1086/173821}

\bibitem[{{Min} {et~al.}(2014){Min}, {Matsumoto}, {Kim}, {Hirota}, {Shibata},
  {Cho}, {Shizugami}, \& {Honma}}]{Min14}
{Min}, C., {Matsumoto}, N., {Kim}, M.~K., {et~al.} 2014, \pasj, 66, 38,
  \dodoi{10.1093/pasj/psu003}

\bibitem[{{Nichols} \& {Slavin}(2009)}]{NicholsSlavin09}
{Nichols}, J., \& {Slavin}, J.~D. 2009, \apj, 699, 902,
  \dodoi{10.1088/0004-637X/699/1/902}

\bibitem[{{Patterson} \& {Raymond}(1985)}]{Patterson_1985}
{Patterson}, J., \& {Raymond}, J.~C. 1985, \apj, 292, 550,
  \dodoi{10.1086/163188}

\bibitem[{{Raymond}(1979)}]{Raymond_1979}
{Raymond}, J.~C. 1979, \apjs, 39, 1, \dodoi{10.1086/190562}

\bibitem[{{Schmid} {et~al.}(2017){Schmid}, {Bazzon}, {Milli}, {Roelfsema},
  {Engler}, {Mouillet}, {Lagadec}, {Sissa}, {Sauvage}, {Ginski}, {Baruffolo},
  {Beuzit}, {Boccaletti}, {Bohn}, {Claudi}, {Costille}, {Desidera}, {Dohlen},
  {Dominik}, {Feldt}, {Fusco}, {Gisler}, {Girard}, {Gratton}, {Henning},
  {Hubin}, {Joos}, {Kasper}, {Langlois}, {Pavlov}, {Pragt}, {Puget}, {Quanz},
  {Salasnich}, {Siebenmorgen}, {Stute}, {Suarez}, {Szul{\'a}gyi}, {Thalmann},
  {Turatto}, {Udry}, {Vigan}, \& {Wildi}}]{Schmid_2017}
{Schmid}, H.~M., {Bazzon}, A., {Milli}, J., {et~al.} 2017, \aap, 602, A53,
  \dodoi{10.1051/0004-6361/201629416}

\bibitem[{{Slavin} {et~al.}(2015){Slavin}, {Dwek}, \& {Jones}}]{Slavin_2015}
{Slavin}, J.~D., {Dwek}, E., \& {Jones}, A.~P. 2015, \apj, 803, 7,
  \dodoi{10.1088/0004-637X/803/1/7}

\bibitem[{{Solf}(1992)}]{Solf_1992}
{Solf}, J. 1992, \aap, 257, 228

\bibitem[{{Toal{\'a}} {et~al.}(2022){Toal{\'a}}, {Sabin}, {Guerrero},
  {Ramos-Larios}, \& {Chu}}]{Toala_2022}
{Toal{\'a}}, J.~A., {Sabin}, L., {Guerrero}, M.~A., {Ramos-Larios}, G., \&
  {Chu}, Y.-H. 2022, \apjl, 927, L20, \dodoi{10.3847/2041-8213/ac589d}

\end{thebibliography}
\bibliographystyle{aasjournal}



\end{document}